\documentclass[usenatbib,useAMS]{mn2e}
\usepackage{epsfig}
\usepackage{times}
\usepackage{amsmath}

\usepackage{verbatim}

\newcommand{\simgt}%
        {\,\hbox{\lower0.6ex\hbox{$\sim$}\llap{\raise0.6ex\hbox{$>$}}}\,}

\title[]{On the $\gamma$-ray emission of Type~Ia Supernovae}
\author[Sim \& Mazzali]{S. A. Sim$^1$, P. A. Mazzali$^{1,2,3}$\\
$^{1}$Max-Planck-Institut f\"{u}r Astrophysik, Karl-Schwarzschildstr. 1,
85748 Garching, Germany\\
$^{2}$Istituto Nazionale di Astrofisica-Osservatorio Astronomico di Trieste, 
Via Tiepolo 11, I-34131 Trieste, Italy\\
$^{3}$Kavli Institute for Theoretical Physics, University of California, Santa Barbara, CA 93106\\
}

\date{25 Oct 2007}

\pagerange{\pageref{firstpage}--\pageref{lastpage}}
\pubyear{}
\begin{document}
\maketitle
\label{firstpage}

\begin{abstract}
A multi-dimension, time-dependent 
Monte Carlo code is used to compute sample
$\gamma$-ray spectra to explore whether unambiguous
constraints could be obtained from $\gamma$-ray observations of type~Ia
supernovae.
Both spherical and aspherical geometries are considered 
and it is
shown that moderate departures from sphericity
can produce viewing-angle effects that are at least as significant as
those caused by the 
variation of key parameters in one-dimensional models. Thus
$\gamma$-ray data could in principle carry some geometrical
information, and caution should be
applied when discussing the value of $\gamma$-ray data
based only on one-dimensional explosion models.
In light of the limited sensitivity of current $\gamma$-ray observatories, 
the computed theoretical spectra are studied to
revisit the issue of whether useful constraints
could be obtained for moderately nearby objects. The
most useful $\gamma$-ray measurements are likely to be of the light
curve and time-dependent hardness ratios, but sensitivity
higher than currently available,
particularly at relatively hard energies ($\sim 2$ -- $3$~MeV), is desirable. 
\end{abstract}

\begin{keywords}
radiative transfer --  methods: numerical -- supernovae: general
\end{keywords}

\section{Introduction}
\label{sect:intro}

Although the paradigm that 
Type Ia supernovae (SNe~Ia) result from the explosions of 
carbon-oxygen white dwarfs is well established,
many issues regarding the nature of the
progenitors and the explosion mechanism remain unclear (see e.g. 
\citealt{hillebrandt00}). Achieving a clearer understanding of
SNe~Ia is important 
because of their role in the chemical
evolution of galaxies and as cosmological distance indicators.

Considerable theoretical effort has gone into
modelling of SNe~Ia explosions. Recently, 
fully three-dimensional (3D) modelling of the explosion
\citep{reinecke02,gamezo03,roepke05,roepke06,jordan07} has
become feasible,
allowing a detailed treatment of the hydrodynamical 
instabilities and turbulence
which play a pivotal role.
The 3D
structure predicted by these models has been shown to affect
observables such as optical/ultra-violet/infrared ({\sc
uvoir}) light curves and spectra 
(e.g. \citealt{kasen06c, sim07, sim07b}).

Although they are primarily detected through their optical emission,
SNe~Ia are
also expected to be  
$\gamma$-ray sources owing to the large masses of radioactive
isotopes synthesised in the explosion 
(e.g. \citealt{travaglio04}). In principle, measurements of
$\gamma$-ray emission from SNe~Ia could provide important diagnostics
since $\gamma$ rays trace almost directly
the mass and velocity distribution of the products of
nuclear burning. Therefore, there has been considerable
theoretical work on SN~Ia $\gamma$-ray emission
(e.g. \citealt{ambwani88, burrows90, mueller91, hoeflich92,
kumagai97, hoeflich98, gomez98, milne04}). Unfortunately, owing to the 
low sensitivity of $\gamma$-ray observatories, 
to date only one SN~Ia (SN1991T)
has been detected (see discussion by
\citealt{milne04}). 
However, with current instrumentation 
such as the 
{\it SPI} \citep{vedrenne03}
on-board {\it Integral} \citep{winkler03}, 
detection of SN~Ia within several Mpc
would be possible
(\citealt{gomez98}). Moreover, increased sensitivity in
future missions should make detection feasible for more distant SNe~Ia.

Most studies of SN~Ia $\gamma$-ray emission focus
on making predictions from individual
models representing specific explosion mechanisms. 
In general, these studies have been restricted to one-dimension
(although see \citealt{hoeflich02}) and to Chandrasekhar-mass
models
(some sub-Chandrasekhar models have been considered, e.g. 
\citealt{hoeflich98}; \citealt{gomez98}).
Such studies demonstrated
that, for nearby SN~Ia, the prospects of obtaining 
useful data are fairly good.
Here we adopt a different but 
complementary approach, motivated by the increasing variety of 
explosion conditions suggested (both in theoretical and semi-empirical
studies): 
rather than determining 
the quality of $\gamma$-ray data
that would be needed to distinguish specific models, we attempt to
show what might be unambiguously determined 
solely from data and the physics of
$\gamma$-ray radiation transport.

We use a multi-dimensional, time-dependent
code to compute $\gamma$-ray spectra for a set of parameterised
geometries
spanning a
broad range in the relevant physical conditions. Using these reference
spectra, we 
highlight
the quantities that would be most 
useful diagnostics
and therefore most worthy of consideration in the design of future 
instrumentation.

In Section~\ref{sect:code}, we briefly describe the code used to compute the 
reference $\gamma$-ray spectra. 
The reference models we employ are
motivated in Section~\ref{sect:models} and their spectra
are discussed
in Section~\ref{sect:spectra}. 
Guided by our reference spectra, in
Section~\ref{sect:diagnostics}
we
examine observational diagnostics.
Finally, in
Section~\ref{sect:summ}, we summarise our results. 

\section{Radiative transfer calculations}
\label{sect:code}

Our calculations were
performed using Monte Carlo methods to track the emission,
propagation, scattering and absorption of quanta which represent bundles
of $\gamma$-ray photons.
Monte Carlo methods are widely used in
modelling $\gamma$-ray spectra for SN ejecta, mostly owing to the
ease with which Compton scattering may be treated in a Monte Carlo 
scheme (see e.g. \citealt{milne04} for discussion 
of SN~Ia $\gamma$-ray transport or \citealt{pozdniakov83} for
an extensive review of Comptonization). They
are also highly effective for 
following radiation transport in SNe~Ia beyond the $\gamma$-regime, 
thereby obtaining bolometric light curves
(following
\citealt{cappellaro97}) and even time-series spectra
\citep{mazzali93,kasen06a}.

The code
used here (\citealt{sim07}) is
closely based on the scheme and test code developed by
\cite{lucy05}.
We briefly summarise its operation 
and describe the physical processes included.
We then describe how the histories of the quanta are used to deduce volume
based emissivities for computation of the emergent
$\gamma$-ray spectra.

\subsection{Overview of code operation}

As described by \citet{lucy05}, the Monte Carlo quanta used in our
code represent indivisible parcels of energy. They begin
as pellets of radioactive material which were 
synthesised during nucleosynthesis in the first few seconds of
the SN explosion.

These pellets decay according to the radioactive 
nuclei they represent
and are converted into $\gamma$-ray packets. The initial
frequencies of these $\gamma$-rays are chosen by sampling the relative
probabilities of the different emission lines that arise from radioactive decay. 

The propagation of the $\gamma$-rays is
followed in 3D space, time and photon frequency until
they either escape from the SN or are lost
from the $\gamma$-ray regime; the latter can occur as a
result of energy loss to non-thermal electrons in Compton
scattering or by photoabsorption. During the
Monte Carlo simulation, the trajectories of the quanta are used to
compute volume-based emissivities; the final spectra are obtained from
a formal solution of the radiative transfer equation using 
these emissivities (see Section~{\ref{sect:emiss}}).

Typically, $4 \times 10^7$ quanta are used in each Monte Carlo
simulation. They are tracked on a homologously expanding $100^3$ Cartesian
grid. 
The simulations cover the time
interval 5 -- 150 dy which is discretised into 50 logarithmically spaced
steps.

\subsection{Radioactive emission}

In the calculations, only $^{56}$Ni and $^{56}$Co are
included as sources of $\gamma$-ray
emission. The relevant nuclear data are
taken from \cite{ambwani88}; their Table 1. All
the $^{56}$Ni is assumed to be produced 
in the first seconds of the explosion
(i.e. at $t = 0$, for our purposes); 
$^{56}$Co is produced only as a result of the subsequent decay of
$^{56}$Ni (i.e. no $^{56}$Co is directly synthesised).

Following \citet{ambwani88} and \citet{lucy05}, positrons released by 
$^{56}$Co decays are assumed to decelerate rapidly and then
annihilate in situ, leading to the emission of two 0.511~MeV
$\gamma$-ray photons in the local comoving frame (i.e. we adopt a
positronium fraction of zero; see e.g. \citealt{milne04} for a discussion). 
The kinetic energy
deposited in the deceleration is assumed to be radiated at much softer
energies and not to contribute to the $\gamma$-ray spectrum.

Although other radioactive isotopes are synthesised in the early
stage of a SN explosion (e.g. $^{44}$Ti, $^{57}$Ni and $^{57}$Co),
they are
expected to have much smaller mass fractions than 
$^{56}$Ni (see e.g. \citealt{travaglio04}) and can be neglected
at relatively early times (i.e. when the light curves are bright and
dominated by $^{56}$Ni and $^{56}$Co).
All other sources of $\gamma$-ray emission (e.g.
nuclear excitation or spallation) are neglected because, 
as discussed by \cite{gomez98},
their contributions are
negligible until very late times.

\subsection{Scattering}

The most important physical process 
the $\gamma$-ray quanta undergo is Compton scattering.
It is treated via the indivisible-energy-packet scheme
presented by \citet{lucy05}.
The Compton cross-section ($\sigma_C$) is given as a function of photon energy
and scattering angle by the Klein-Nishina formula. 
In all calculations, we 
neglect the 
initial electron momentum in
the comoving frame.

\subsection{Absorption}

Two $\gamma$-ray destruction processes are included in the
calculations.
First, photoabsorption is included following
\citet{lucy05}. The photoabsorption cross-section per unit gram
($\kappa_{ph}(E)$)
is allowed to depend
on composition in a simple manner:

\begin{equation}
\kappa_{ph}(E) = X_{\mbox{\scriptsize Fe-grp}} \; \kappa_{28}(E) + (1 -
X_{\mbox{\scriptsize Fe-grp}}) \;
\kappa_{14}(E)
\label{eqn:sigmaph}
\end{equation}
where $X_{\mbox{\scriptsize Fe-grp}}$ is the local mass fraction of iron-group
elements and 
$\kappa_{28}(E)$ and $\kappa_{14}(E)$ are the photoabsorption
cross-sections per gram typical of high-and intermediate-mass elements,
respectively (the subscripts refer to the ``typical'' atomic mass
number, $Z$). 
Specific expressions for $\kappa_{28}(E)$ and
$\kappa_{14}(E)$
are 
obtained from \citet{veigele73} following
\citet{ambwani88}.

Second, for $\gamma$-rays above 1.022~MeV, pair
production is included. The cross-section for this process, 
$\kappa_{pp}(E)$,
is allowed
to depend on composition in the same way as photoabsorption
(Equation~\ref{eqn:sigmaph}).
The representative cross-sections for intermediate- and
high-mass elements ($Z=14$ and $28$) are obtained using equation~2
of \citep{ambwani88}.

When a pair production event occurs, probabilities are assigned to
determine whether the 
outcome is
a positron or a non-thermal-electron packet (see
\citealt{lucy05}). The kinetic energy of both 
positron and electron are assumed to be lost from the
$\gamma$-ray regime. However, as with the positrons emitted by
$^{56}$Co, the positrons resulting from pair-creation are assumed to
annihilate in situ, leading to the emission of two $\gamma$-rays at
0.511~MeV.

\subsection{Extraction of spectra}
\label{sect:emiss}

As discussed by \citet{lucy99,lucy03,lucy05}, 
and in the context of the
current code by \citet{sim07}, optimal use of a
Monte Carlo simulation of radiation transport is generally obtained by
recording volume based estimators and using these to extract the
quantities of interest. The
estimators required are those from which we obtain
the Compton emissivity ($\eta^C$, a function of position, time, frequency and
observer line-of-sight direction)
and the 0.511~MeV-emissivity arising from pair-production ($\eta^{pp}$). With these
emissivities known, a formal solution of the radiative transfer
equation can be performed to deduce the spectrum (see \citealt{lucy99,
sim07}). The third emissivity term required for the formal solution --
describing direct $\gamma$-ray emission from
radioactive decay -- does not require a Monte Carlo estimator since it
can be derived from the assumed distribution of radioactive
material.

Monte Carlo estimators for $\eta^C$ are recorded
in every grid cell during all relevant time steps in the simulation,
on a pre-determined frequency grid.
In the comoving frame

\begin{equation}
\eta^{C}_{\mbox{\scriptsize cmf}} = \frac{n_{e}}{V \; \Delta t \; \Delta \nu}
\sum_{\mbox{\scriptsize $\gamma$-paths}}
\frac{\epsilon_{\mbox{\scriptsize rf}}}{f}
\mbox{d}s
(1 - 2\mbox{\boldmath $v \cdot \hat{n}$}/c)
\left( {\frac{\mbox{d} \sigma}{\mbox{d} \Omega}}
\right)_{\nu_{\mbox{\scriptsize cmf}}, \mu_{\mbox{\scriptsize
cmf}}}
\end{equation}
where the summation runs over all $\gamma$-ray trajectories
inside the grid cell during the time step; 
$\mbox{d} \sigma / \mbox{d} \Omega$ is the
Klein-Nishina differential cross-section, evaluated at the comoving
frequency of the $\gamma$-ray packet ($\nu_{\mbox{\scriptsize cmf}}$)
and for a scattering angle specified by $\mu_{\mbox{\scriptsize
cmf}}$ (the cosine of the comoving-frame angle between the direction of packet
propagation and the observer line-of-sight);
{\boldmath $v$} is the ejecta velocity,
{\boldmath $\hat{n}$} the rest-frame direction in which the packet is 
propagating, $\epsilon_{\mbox{\scriptsize rf}}$ the rest-frame packet
energy and $\mbox{d}s$ the trajectory length; $f$ is the ratio of
photon energies before and after Compton scattering

\begin{equation}
f = 1 + \frac{h \nu_{{\mbox{\scriptsize cmf}}}}{m_{e} c^2} (1 -
\mu_{\mbox{\scriptsize cmf}}) \; \; ;
\end{equation}
$n_{e}$ is the electron number density, 
$V$ is the volume of the grid cell, $\Delta t$ the duration of the
time step and $\Delta \nu$ the width of the frequency bin. The
frequency bin to which a given $\gamma$-ray trajectory is assigned is
determined by $\nu_{\mbox{\scriptsize cmf}} / f$, the
comoving frame frequency hypothetical photons from the $\gamma$-packet
would have were they to
undergo Compton scattering into the observer line-of-sight.

The corresponding estimator for
$\eta^{pp}$
is simpler. Since
the emission is restricted to a single comoving-frame photon energy 
(0.511~MeV), only one estimator per grid cell per time step is
required for this process; it is given by

\begin{equation}
\eta^{pp}_{\mbox{\scriptsize cmf}} = \frac{1}{4 \pi V \; \Delta t}
\sum_{\mbox{\scriptsize $\gamma$-paths}}
{\epsilon_{\mbox{\scriptsize rf}}} \;
\frac{2 m_e c^2}{h \nu_{{\mbox{\scriptsize cmf}}}} \; \mbox{d}s \;
(1 - 2\mbox{\boldmath $v \cdot \hat{n}$}/c) \;
\kappa_{pp} \; \; .
\end{equation}
The summation now runs over all $\gamma$-trajectories in the
cell during the time step and for which the photon energy is above the
pair-creation threshold.
Although $\eta^{C}_{\mbox{\scriptsize cmf}}$ is a specific emissivity,
$\eta^{pp}_{\mbox{\scriptsize cmf}}$ is not -- hence they differ dimensionally
by [Hz$^{-1}$]. Throughout this study it is assumed that emission-line shapes
are determined by the macroscopic velocity field. Thus the intrinsic emission
profile associated with $\eta^{pp}_{\mbox{\scriptsize cmf}}$ is taken to be 
arbitrarily narrow.

All the spectra shown in this paper were computed
via formal solutions of the radiative transfer equation 
using the Monte Carlo emissivity estimators given above.
The accuracy and validity of the use of the estimators was separately
tested by comparing spectra obtained from the formal solution with
those obtained by direct frequency binning of emergent Monte Carlo
quanta for our spherically symmetric ``control'' model (Model SC, see
below). To the level of the Monte Carlo noise in the
frequency-binned spectrum (typically a few
percent), 
the two methods produced indistinguishable results,
as expected.

\section{Models}
\label{sect:models}

\begin{figure}
\epsfig{file=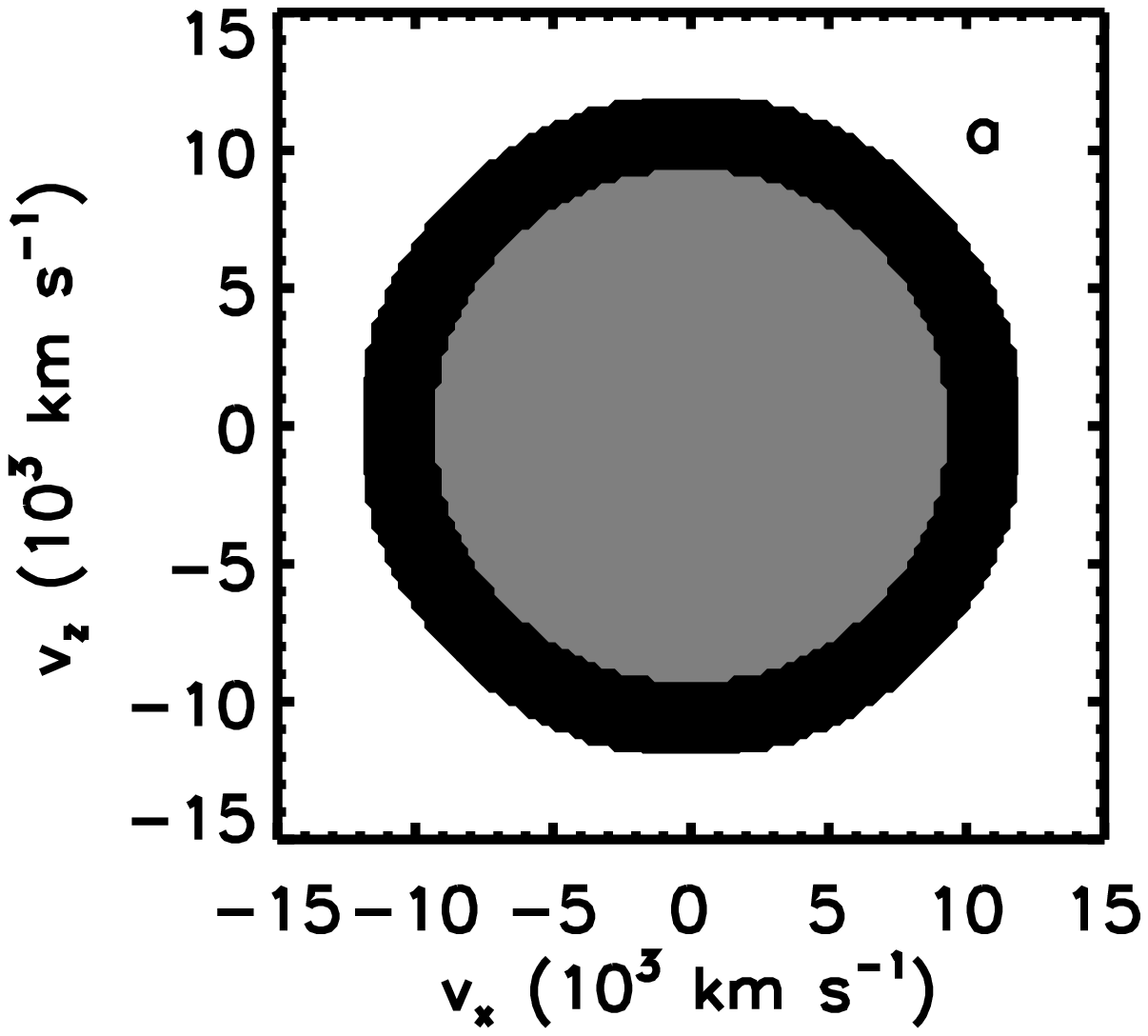, width=4.3cm}
\epsfig{file=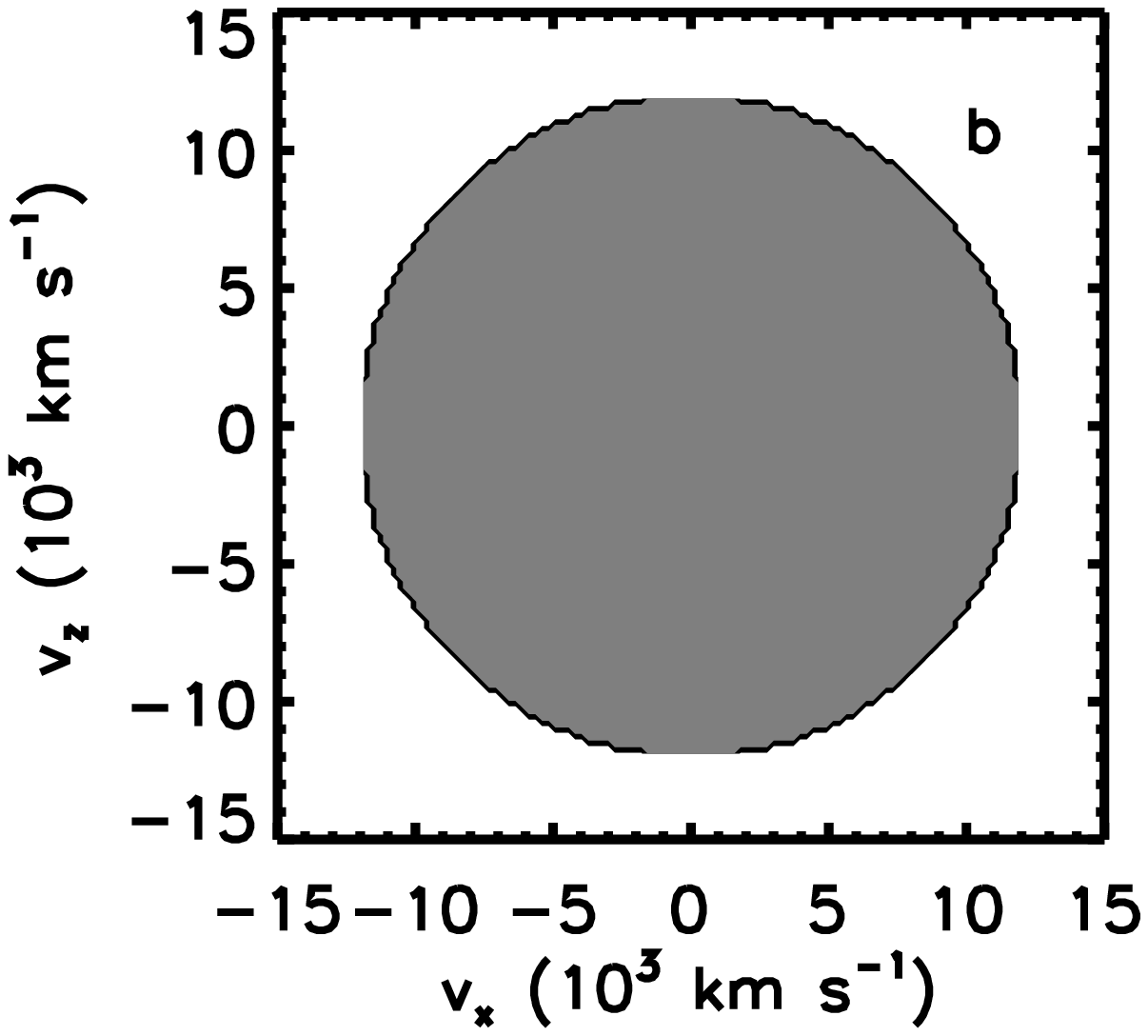, width=4.3cm}\\
\epsfig{file=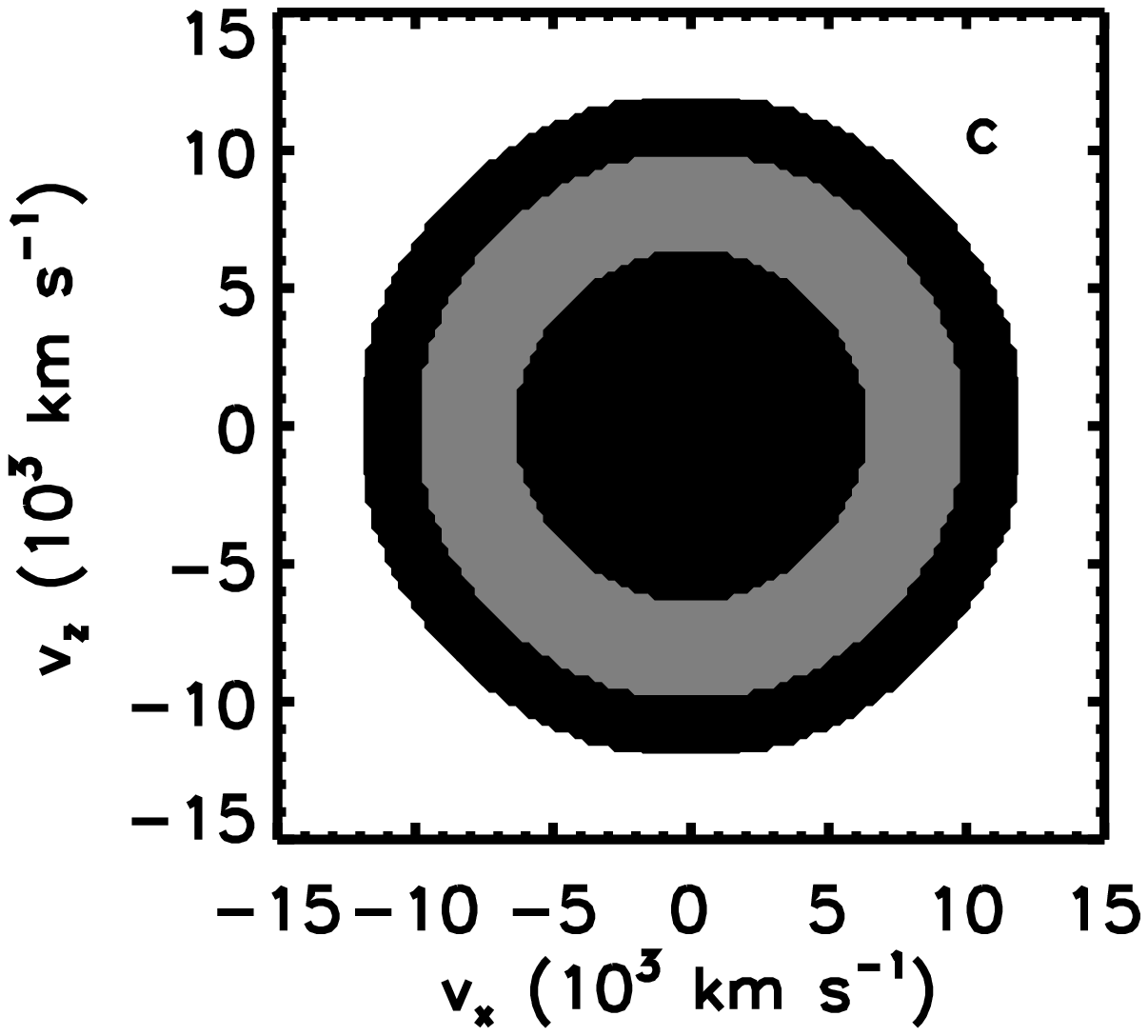, width=4.3cm}
\epsfig{file=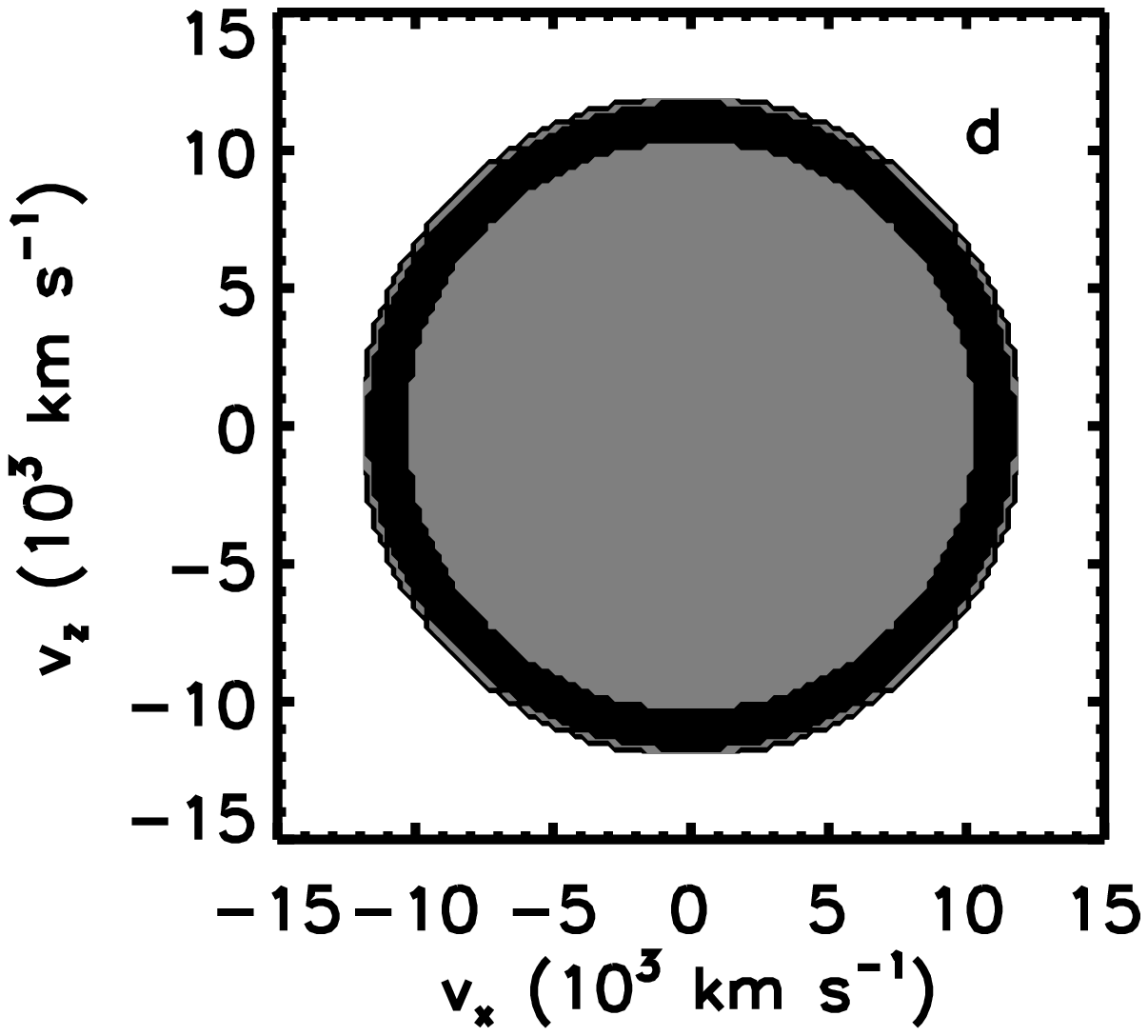, width=4.3cm}\\
\epsfig{file=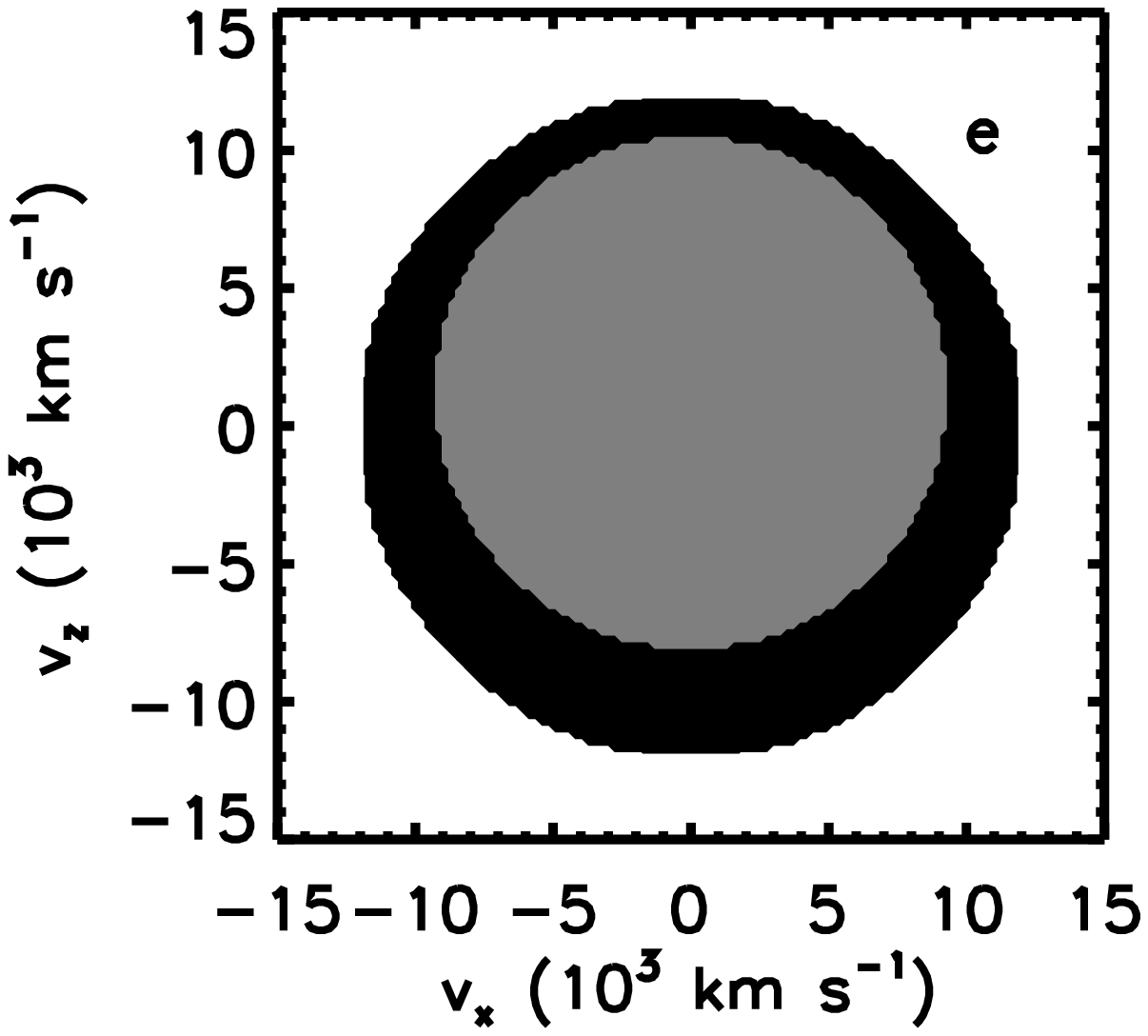, width=4.3cm}
\epsfig{file=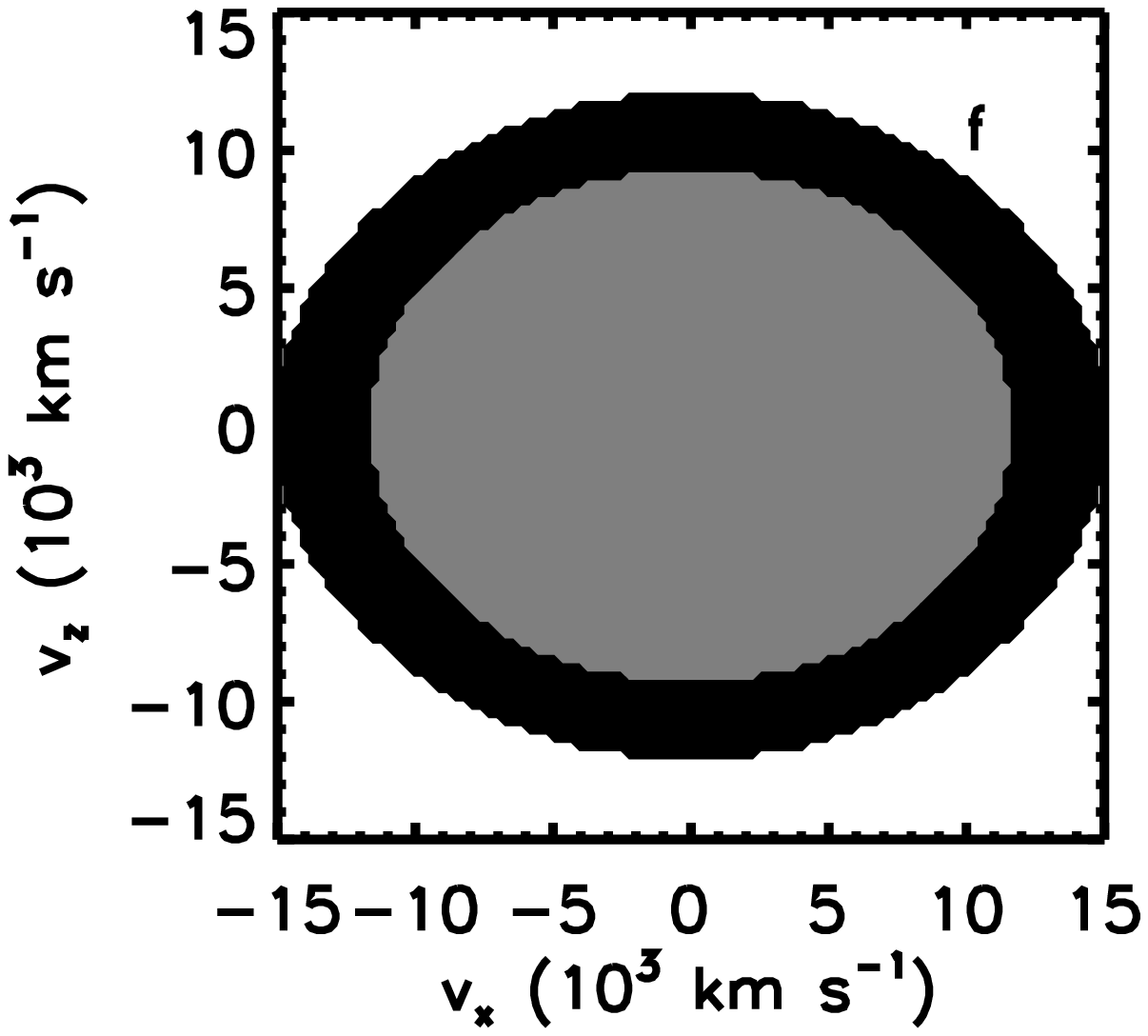, width=4.3cm}
\caption{
Slices through the x-z velocity-space 
plane showing the distribution of $^{56}$Ni adopted in the models. Grey
indicates regions which initially contain $^{56}$Ni while black
shows the volumes that do not.
Panel~a shows the basic geometry used for
Models~SC, SS and SFeR. Panels b, c, d, e and f
show,
respectively, Models~SM, SFeC, SNiS, AO and AE. All models are symmetric under
rotation about the $z$-axis.
\label{fig:models}
}
\end{figure}

We computed $\gamma$-ray radiation transport for 
a set of eight models. These are not tied to specific 
progenitor/explosion scenarios in detail but are all rooted in previous
discussions of SNe~Ia.

Although quite diverse, our models are not 
remotely exhaustive of all possible conditions or
geometries. 
However, as stated in Section~\ref{sect:intro}, 
our intention is not to address
specific explosion models but to
explore 
whether
relatively model-independent information can be inferred from
the $\gamma$-ray regime.

Compared
to other spectral regions (e.g. ultraviolet, optical, infrared etc.),
the
$\gamma$-ray spectrum is comparatively simple -- 
its brightness and 
shape are mainly determined by only two quantities: the instantaneous 
number of 
radioactive decays
and the optical depth to Compton scattering ($\tau_C$). 
Thus to first order,
$\gamma$-ray data constrain only
the properties that are relevant to these two quantities. 
Of course, 
other processes (e.g., photoabsorption) have a role in establishing the 
$\gamma$-ray spectrum, but their effects are generally secondary
and they may be disentangled.

The most easily understood parameter for $\gamma$-ray
emission is the total mass of radioactive
material ($M_{\mbox{\scriptsize Ni}}$). 
Except for two cases (see below), we fix this at 
0.6~M$_{\odot}$.
To zeroth-order, a larger
mass leads to brighter emission. 
However, there are 
additional effects.

A second relevant quantity is the 
velocity distribution of $^{56}$Ni since Doppler shifts
are important in setting emission line widths (see
e.g. \citealt{milne04}). 
Detailed
radiation transport is not really needed to explore this
simple relationship, particularly since only
extremely sensitive
observations could constrain line
profiles (see e.g. \citealt{gomez98}).

Radiation transport {\it is} important for
the shape of both the
light curve and 
the spectrum.
The dominant process affecting 
$\gamma$-ray transport in SNe~Ia is Compton scattering.
Neglecting light travel time effects 
and relativistic corrections,
$\tau_C$ is sensitive only to the column density
of target electrons ($N_{e}$). Since the photon energies under consideration
($\sim 1$~MeV) are very large compared to the binding energies of
electrons in atoms (typically in the 10~eV -- 10~keV range), it is
reasonable to regard all electrons as available Compton-targets. Thus
the Compton optical depth is independent of the thermal, ionization
and excitation
states of the plasma and, to a good approximation, 
also of the density profile along the
line-of-sight since both the velocity and the velocity gradient
appear only through Doppler factors. This
is in contrast to other wavebands (e.g. the
optical) where line opacity dominates giving greater significance to 
the velocity structure.

In light of these considerations, our models explore a wide
range of distributions of radioactive source material with $\tau_C$ 
-- as a minimal standard of data quality, 
this distribution should be
constrained before data can 
support or exclude any specific scenario.

Six of our models are spherically
symmetric while the other two are aspherical. 
As mentioned in Section~\ref{sect:intro},
most previous
studies of $\gamma$-ray spectra were restricted to 1D
calculations. 
In one of the few
multi-dimensional studies of SN~Ia $\gamma$-ray spectra, 
\citet{hoeflich02b} considered an
elliptical SN explosion and showed that significant viewing angle
effects may occur, particularly at relatively early times. 
That result is comparable to those from studies of aspherical
SNe~Ia in other wavebands (e.g. 
\citealt{hoeflich91,kasen06c,sim07b, hillebrandt07})
which show that
viewing angle effects are potentially 
observable and could be confused with other phenomena. Therefore,
one of our objectives is to add aspherical models to the zoo of
other possibilities and consider whether they could be
distinguished.

The eight models we use are described below.
They are labelled by two to 
four letters -- the first letter identifies whether the 
model is spherical (``S'') or aspherical (``A'') while the 
others abbreviate
the model properties.
The model geometries are shown in 
Fig.~\ref{fig:models}. 
All models are assumed to be in
homologous expansion and, for simplicity, to have uniform mass-density
(as noted above, the density profile is relatively unimportant for the
synthesis of the $\gamma$-ray spectrum).

\subsection{Model SC: control model}

Our first model, which we use as the standard of
comparison for all the others, is spherically symmetric with a total
mass $M_{\mbox{\scriptsize T}} = 1.4$~M$_{\odot}$ and 
$M_{\mbox{\scriptsize Ni}} = 0.6$~M$_{\odot}$. The maximum ejecta velocity is set at
$v_{\mbox{\scriptsize max}} = 1.2 \times 10^{9}$ cm~s$^{-1}$. 
This is lower than the maximum velocities in real SNe but
representative of the velocity regime in which the density is large 
and is therefore appropriate given our choice of a uniform density profile 
(see above).
In this model, the $^{56}$Ni is located in a spherical region
at the centre of the ejecta. The volume of this region is fixed
by an adopted $^{56}$Ni mass fraction of 0.9.
The region outside the central concentration of Ni is assumed to be
composed of intermediate mass elements.

\subsection{Model SS: super-Chandrasekhar mass model}

Model SS has the same geometry and Ni-distribution as Model~SC but its
densities are everywhere higher by a factor of 1.5. Thus, this
model is ``super-Chandrasekhar'', having
$M_{\mbox{\scriptsize T}}=2.1$~M$_{\odot}$ and $M_{\mbox{\scriptsize Ni}} = 0.9$~M$_{\odot}$.
This model explores $\gamma$-ray emission from
objects in which $\tau_C$ is higher than in Model~SC. Although outside the standard SN~Ia paradigm, 
Super-Chandrasekhar explosions have been proposed for
unusually bright events such as SN~1991T \citep{fisher99}, 
SN~2003fg \citep{howell06} and SN~2006gz \citep{hicken07}.

\subsection{Model SFeR: Fe-rich ejecta}

This model isolates the effect of composition. 
It adopts the same geometry, mass and
Ni-distribution as Model~SC. However, it assumes that
the majority of the material that is not initially $^{56}$Ni consists of
other heavy nuclei; the photoabsorption and pair-production
cross-sections are thus different from Model~SC.

\subsection{Model SM: well-mixed model}

This model is identical to Model~SC ($M_{\mbox{\scriptsize
T}} = 1.4$~M$_{\odot}$, $M_{\mbox{\scriptsize Ni}} = 0.6$~M$_{\odot}$)
except that $^{56}$Ni is
distributed uniformly throughout the volume of the ejecta (see
panel~b of Fig.~\ref{fig:models}). 

\subsection{Model SFeC: stable Fe core}

Recently, \citet{mazzali07} have argued for  
a model which may 
account for most SNe~Ia. Analysing spectra of a large sample of SNe, they
concluded that the characteristic structure consists of a low-velocity core of
stable iron-group material surrounded by a $^{56}$Ni-rich
region. Outside this region, the material underwent incomplete nuclear 
burning and is dominated by intermediate-mass nuclei.
Motivated by their study, we include a model based on such a structure 
(the geometry is shown in panel~c of  Fig.~\ref{fig:models}).
The inner stable iron-group core has a mass of
0.2~M$_{\odot}$. 
A pure $^{56}$Ni-region lies 
immediately above this core and contains 
0.6~M$_{\odot}$. The outermost region is 
composed of intermediate-mass elements only.

\subsection{Model SNiS: Ni-rich core and surface layer}

This model differs from all others as it includes two
physically separated $^{56}$Ni-rich regions: a massive core and a thin
surface layer. 
This geometry 
provides an alternative  
to the 
``super-Chandrasekhar'' scenario (see Section 3.2)
for 
unusually bright events.
For this model we retain $M_{\mbox{\scriptsize T}} = 1.4$~M$_{\odot}$, 
adopt a core $^{56}$Ni-mass of 0.9~M$_{\odot}$ and assume that the
outermost 0.1~M$_{\odot}$ of the ejecta is also pure $^{56}$Ni. 
This outer layer is motivated by the early-time spectroscopic
behaviour of SN1991T \citep{mazzali95}.
The material 
sandwiched between the core and the outer Ni-rich shell is taken to be 
composed of intermediate mass elements.
This model has the largest $M_{\mbox{\scriptsize Ni}}$ in our study, as required to 
account for the brightness of events such as SN1991T.

\subsection{Model AO: off-set Ni}

Here, the basic model is identical to Model~SC but the
centre-of-mass of the Ni ball is displaced along the $z$-axis 
by 10 per cent of the outer ejecta radius. 
This leads to an aspherical model, very similar to
those discussed by 
\cite{sim07b}.

\subsection{Model AE: ellipsoidal ejecta}

In this model, the maximum ejecta velocity varies with direction,
producing an ellipsoidal SN. 
This is the same basic geometry as considered by \cite{hoeflich02b}. 
Ellipsoidal explosions have been suggested following
detection of polarisation in SNe~Ia (e.g. \citealt{howell01, wang03}).
The maximum velocity ( $1.5 \times 10^{9}$
cm~s$^{-1}$)
occurs at the equator. The axis ratio is set to 5:4; therefore the terminal
velocity of the ejecta in the polar direction is $1.2 \times 10^{9}$
cm~s$^{-1}$.
The Ni-distribution is ellipsoidal in the same sense as the total mass
distribution and remains centrally concentrated.
As in all the other models, the mass-density remains uniform.

\begin{figure}
\epsfig{file=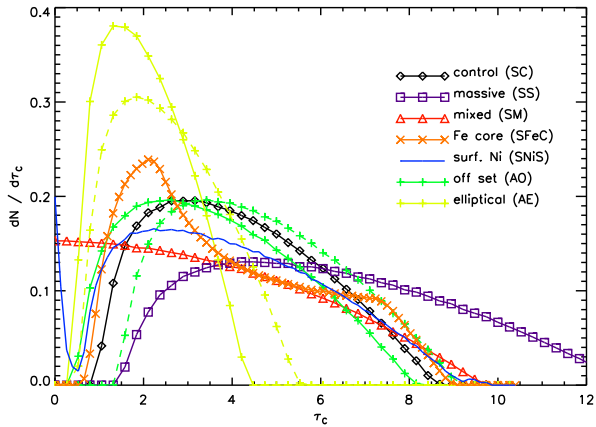, width=\columnwidth}
\caption{
Distribution of source $^{56}$Ni with 
electron scattering optical depth (computed in
the Thompson limit) for the models. 
The distribution is normalised to $\int_0^\infty \frac{\mbox{d}
N}{\mbox{d} \tau_C} \; \mbox{d} \tau_C = 1$.
The symbols
are indicated in the figure. Model~SFeR is not plotted since,
with respect to this distribution, it is identical to Model~SC. For
both the aspherical models (Models AO and AE), two curves are plotted 
-- these represent the
extremes with respect to viewing angle: for Model AO, the solid line is
for a line-of-sight parallel to the direction in which the Ni-blob is
displaced from the centre of the SN while the dashed line is for the 
anti-parallel light-of-sight; for Model AE, the solid line is for
viewing down the short axis of the ellipsoid while the dashed line is
for a line-of-sight along the long axis.
}
\label{fig:taus}
\end{figure}

\subsection{Distribution of emission with opacity}

As discussed above, since the $\gamma$-ray spectrum is primarily
determined by Compton scattering, we expect that
a highly relevant quantity is the distribution of $^{56}$Ni with
$\tau_C$
(see Fig.~\ref{fig:taus}). 
$\tau_C$ is a decreasing 
function of both time and photon energy. However, since the energy
dependence is universal and, in an homologous
flow, the time dependence is very simple ($\tau_C \propto t^{-2}$),
it is sufficient to compare the distribution for
one photon energy at one time. For convenience, the optical depths
used in Fig.~\ref{fig:taus} are computed using the Thompson limit to $\sigma_C$
at time $t = 50$~dy.

Fig.~\ref{fig:taus} shows that the models considered
{\it are} quite diverse and reasonably complete, covering a range of 
plausible single-peaked distributions. At the
epoch considered, the radioactive material in Model~SC is concentrated at
moderate optical depths, and its distribution is bracketed by the
extremes shown in Model~AO. 
Models~SFeC and AE have lower optical depths, but their distributions are still peaked
significantly away from $\tau_C = 0$. Model~SM has an extreme
distribution, with a significant fraction of the source material at
very low optical depths ($\tau_C \sim 0$). At the opposite limit,
Model~SS has higher opacities, resulting from the larger density.
Model~SNiS differs from the others in having two peaks in its distribution -- most of the 
Ni lies at moderate optical depths but the thin surface layer provides a distinct peak at 
very low optical depth.

Model~SFeR is not shown in Fig.~\ref{fig:taus}; its
distribution of $^{56}$Ni with $\tau_C$ is 
identical to Model~SC so that it can isolate
the signatures of photoabsorption. 

\section{Discussion of spectra}
\label{sect:spectra}

\begin{figure}
\epsfig{file=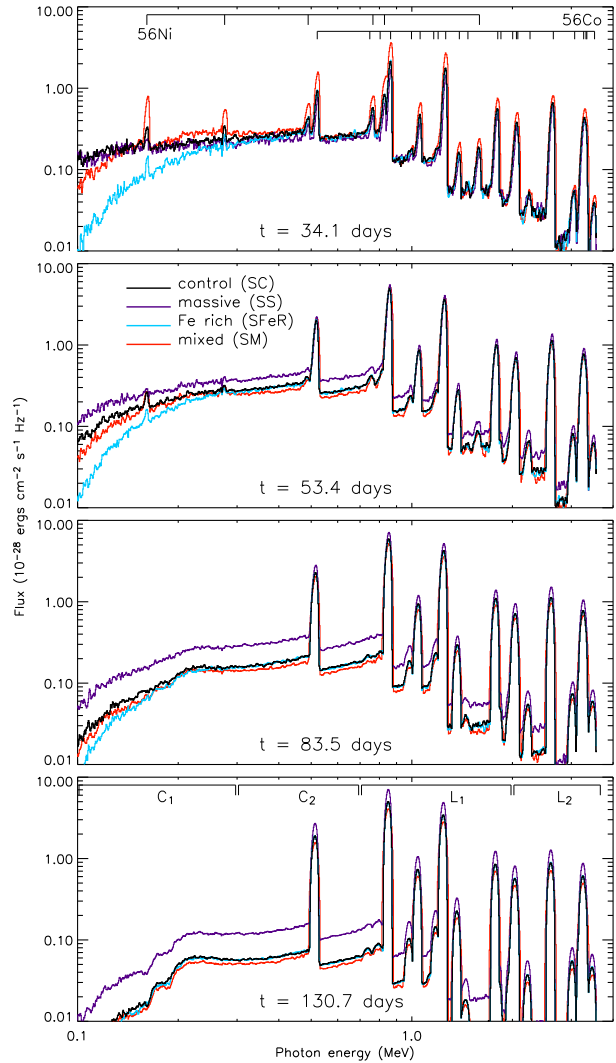, width=8.5cm}
\caption{
Representative
time series of $\gamma$-ray spectra computed with four spherically
symmetric models (identified in the second panel) at
times after explosion as indicated on each panel. 
The $^{56}$Ni and $^{56}$Co emission lines are identified in the top panel. 
The four flux-bands 
discussed in Section 5.3 are indicated in the bottom panel.
Fluxes are for
a source distance of 1~Mpc.
\label{fig:spec}
}
\end{figure}

Fig.~\ref{fig:spec} shows a time series of spectra computed from
Model~SC and spectra from three other spherical
models (SS, SFeR and SM) are over-plotted. 
(For clarity, spectra from the remaining four models 
are not plotted. They are, however, included in all the discussions 
of observable diagnostics in subsequent sections.)

At times close to maximum light the $\gamma$-ray spectrum
consists of strong emission
lines, mainly due to $^{56}$Co, with significant continuum arising
from Compton scattering of line photons. Since optical depths
decrease with time in the expanding ejecta, the strength of
the lines relative to the continuum increases with time. 
At early times, 
the ``mixed'' model
(Model~SM) shows a harder continuum than the standard model
and stronger Ni emission lines  
(particularly 0.158~MeV and 0.275~MeV). These lines are
also fairly strong in Model~SNiS (not plotted) where they originate in the surface layer of $^{56}$Ni.
As noted by
\citet{gomez98}, these Ni
lines can only form when the source Ni lies at
small $\tau_C$; since
$\sigma_C$ decreases with increasing photon energy,
the soft-energy lines are degraded most easily, becoming
swamped by photons down-scattered from harder energies.

In Model~SS, the 0.158 and 0.275~MeV lines are almost completely
buried by the strong continuum which
persists until well after maximum light as a consequence of the
high $\tau_C$'s in this model.

Model~SFeR 
shows the effect of
composition as intended. Above about 0.3~MeV, its
spectra are indistinguishable from those of Model~SC. At
soft energies, 
however, 
the Model~SFeR flux is lower by
up to an order of magnitude. This is
due entirely to the difference in the photoabsorption cross-section
arising from the choice of composition. Although weaker,
a similar effect
appears in Model~SNiS because of the relatively large amounts
of iron-group material in that model, particularly in the surface layer.

Should
energy-resolved data of sufficient sensitivity to
measure both the line and continuum emission across the entire 
0.1 -- 3.5~MeV spectrum be obtained, 
it would be used to evaluate SN models by direct comparison. 
However, given the rarity of very nearby SNe~Ia and
the limited sensitivity of 
$\gamma$-ray observatories, in the next section 
we will consider what
information could be extracted from low quality $\gamma$-ray data in
the form of simple diagnostics. 
We will focus on line and continuum fluxes. 
Although the energy resolution of $\gamma$-ray instruments can be
high enough to resolve spectral lines (e.g. with 
 {\it SPI Integral}, see \citealt{roques03}),
sensitivity limits make it practically
impossible to measure detailed line shapes except for extremely nearby
events \citep{gomez98}. 

\section{Observational diagnostics}
\label{sect:diagnostics}

We consider three different potential diagnostics
that could, in princple, be obtained from $\gamma$-ray data.
Throughout our discussion, we use theoretical source count rates to
illustrate the value of the diagnostics for distinguishing our
models. We do not, however, consider any additional effects
(e.g. instrumental noise) which would make detections much harder with
current observatories (see e.g. \citealt{gomez98}).
 
We begin with the
simplest diagnostic, the energy-integrated light curve.

\subsection{$\gamma$-ray light curves}

\subsubsection{Discussion}

The $\gamma$-ray light curve is the
most useful for constraining 
$M_{\mbox{\scriptsize Ni}}$
since in this spectral
region the decay products of 
$^{56}$Ni and its daughter nucleus $^{56}$Co are seen most directly.

\begin{figure}
\epsfig{file=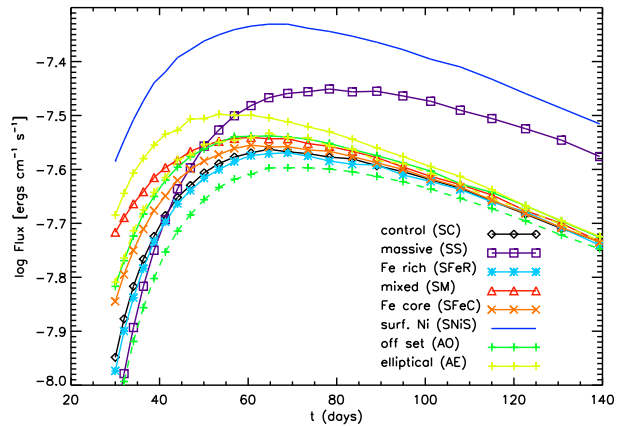, width=\columnwidth}
\caption{
Model light curves for the 0.1 -- 3.5~MeV $\gamma$-ray region.
The flux is given for a source distance of 1~Mpc.
Times are measured from the explosion.
For
the aspherical models (AO and AE), two curves are plotted 
-- as in Fig.~\ref{fig:taus}, these represent the
extremes with respect to viewing angle.
\label{fig:tot_lcs}
}
\end{figure}

Fig.~\ref{fig:tot_lcs} shows energy-integrated (0.1 -- 3.5~MeV)
light curves from the models. All the light curves
have a characteristic, single-peaked shape. They 
peak later ($t_{\mbox{\scriptsize peak}} = 50$ -- 80~dy) and 
are simpler in
shape than model bolometric (or optical/infrared) light curves (see
e.g. \citealt{blinnikov06,kasen06a,kasen06b,kasen07}) since photon
trapping plays a far less significant role in the $\gamma$-ray regime.

After about 100~dy, all our models are fairly optically thin and
the $\gamma$-ray luminosity is an unambiguous measure
of $M_{\mbox{\scriptsize Ni}}$ via

\begin{equation}
\begin{array}{l l l}
  L_{\gamma}(t \simgt 100~dy) & \approx & \epsilon_{\mbox{\scriptsize Co}} M_{\mbox{\scriptsize Ni}}
\exp(-t / t_{\mbox{\scriptsize Co}}) / {t_{\mbox{\scriptsize Co}}} \\
 & \approx & 1.23 \times 10^{43} \frac{M_{\mbox{\scriptsize Ni}}}{M_{\odot}} \exp(-t / t_{\mbox{\scriptsize Co}})
\mbox{~~ergs s$^{-1}$}
\end{array}
\end{equation}
where $L_{\gamma}(t)$ is the emitted $\gamma$-ray luminosity, $\epsilon_{\mbox{\scriptsize Co}}$ is the total
$\gamma$-ray energy released during the decay of one gram 
of $^{56}$Co and $t_{\mbox{\scriptsize Co}}$ is the
$^{56}$Co lifetime.
In Fig.~\ref{fig:tot_lcs}, this optically-thin behaviour causes the
light curves of all models with the same adopted $M_{\mbox{\scriptsize
Ni}}$ (SC, SFeR, SM, SFeC, AO and AE) to converge for $t >
100$~dy while models with higher 
$M_{\mbox{\scriptsize Ni}}$ (SS and SNiS) are
proportionally brighter.

At earlier times, individual model light curves differ
significantly -- the spread is about a factor of two
at $t \sim 40$~dy. 
The light curves of models 
containing Ni at
relatively low $\tau_C$ (SM, AE, SNiS, SFeC and face-on AO; see
Fig.~\ref{fig:taus}) tend to rise quickly and
peak early but the light curve shapes are different
and do not adhere to a one-parameter family. The
aspherical models show viewing angle dependence of up to 50 per cent at
early times, decreasing to about 10 -- 20 per cent at
$t_{\mbox{\scriptsize peak}}$
which is comparable
to differences between our various spherically symmetric models.

The integrated light curve is almost insensitivity to
photoabsorption (in Fig.~\ref{fig:tot_lcs}, compare Models~SFeR
and SC) -- most
of the $\gamma$-rays are too hard for
significant photoabsorption to occur, regardless of the composition of
the absorbing material. 

\subsubsection{Diagnostic value}
\label{sect:lc_obs}

With good time coverage (40 -- 120 dy after explosion), 
the $\gamma$-ray light curve alone can provide 
useful information: 
at late times it
gives the most direct measurement of $M_{\mbox{\scriptsize Ni}}$ 
while at earlier stages it is
sensitive to the distribution of $\tau_{C}$
(i.e. electrons and therefore total mass). 

To assess the 
diagnostic value of a SN~Ia $\gamma$-ray light curve, we
converted our theoretical light curves to 
source counts. For illustrative purposes, we used the energy-dependent
effective area of {\it Integral SPI} 
\citep{attie03,diehl03,sturner03}, adopted a distance of 18~Mpc to the SNe (an approximate
distance to the Virgo cluster, \citealt{fouque01}) and assumed an integration 
time of $10^5$~s ($\sim 1$ dy).
Under these conditions Model~SC would give 
$\sim 460$ photon counts 
at $t \sim 50$~dy, decreasing to $\sim 310$ counts
at $t \sim 100$~dy. 

Therefore, if photon counting
statistics were the only limitation, it would be quite feasible to 
deduce SN~Ia $^{56}$Ni-masses to a precision of about 10~per cent for
objects at distances up to Virgo.
Some
rough information on the 
light curve shape between about 40 and 100 dy after the explosion
could
also be obtained.
The level of precision would be comparable to 
the typical differences between the models
investigated here, so that a time sequence of data 
could place fairly robust
constraints not only on the total $^{56}$Ni mass, but also on its
distribution with opacity.

Of course, these estimates are very crude and optimistic since
additional sources of error (e.g. background) have not been
considered, but they do suggest that, in
principle, useful information is contained in light curve data.

\subsection{Line ratios}
\label{sect:line_rat}

\begin{figure*}
\epsfig{file=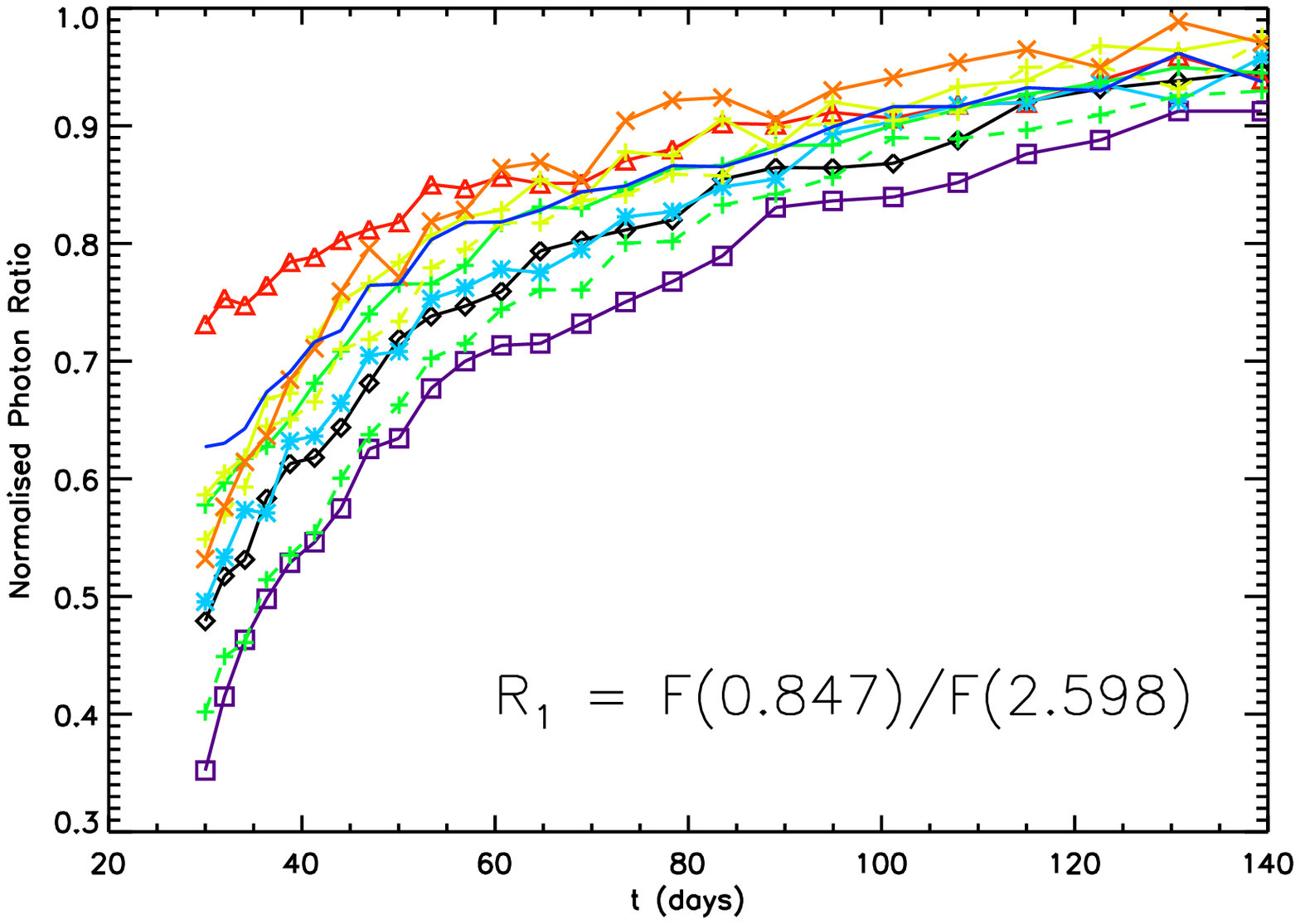, width=5.8cm}
\epsfig{file=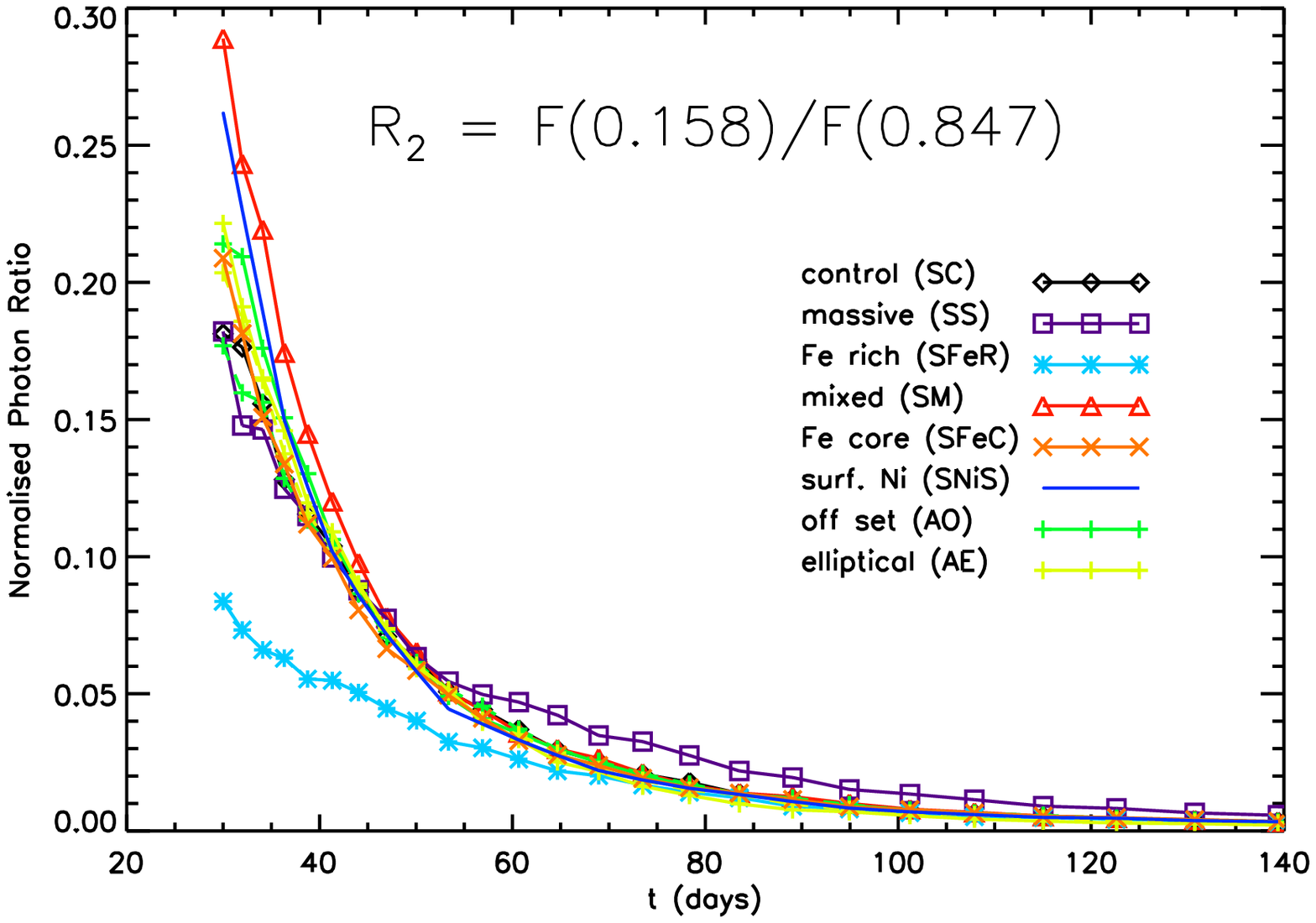, width=5.8cm}\\
\caption{
Peak line flux ratios 
from the various models as functions of time since explosion.
The left panel shows the ratio 
$R_1 = F($0.847 MeV$) / F($2.598 MeV$)$ and the right panel
 $R_2 = F($0.158 MeV$) / F($0.847 MeV$)$.
For
the two aspherical models (Models AO and AE), two curves are plotted 
-- as in Fig.~\ref{fig:taus}, these represent the
extremes with respect to viewing angle.
Flux ratios are normalised to the optically thin limit (thus, the 
normalised $R_{1}$-ratio asymptotes
to 1.0 while $R_2$ asymptotically approaches
the instantaneous mass ratio of $^{56}$Co to $^{56}$Ni).
\label{fig:line_rat}
}
\end{figure*}

\subsubsection{Discussion}

$\gamma$-ray emission line
peak-intensity ratios provide
simple diagnostics 
\citep{burrows90,hoeflich98,gomez98}.
As in earlier studies, we examine only a few lines and their ratios
in detail.
Owing to the simplicity of spectral formation in
the $\gamma$-ray regime, exactly which lines are considered is fairly
unimportant -- the qualitative behaviour of any line ratio will be 
similar to that of one of those we show.

Since the strongest lines
originate in the decay of $^{56}$Co, their intrinsic relative 
strengths are set by 
nuclear physics alone -- thus late-time measurements of line ratios
yield no information that cannot be gleaned from the energy-integrated
light curve.
However, at early times, the energy sensitivity of $\sigma_C$
makes the intensity ratio of lines at
different energies sensitive to $N_{e}$
\citep{hoeflich98}.

The left panel of Fig.~\ref{fig:line_rat} shows 
the ratio
$R_1 = F($0.847 MeV$) / F($2.598 MeV$)$. Both lines in this ratio are 
from the decay of
$^{56}$Co. We chose to consider this particular pair of lines since they are 
strong, unblended and well-separated in energy. Although an even more
widely spaced
pair of lines associated with $^{56}$Co could be achieved considering 
the 0.511~keV positron-annihilation line, we prefer to avoid reliance
on our adopted positronium fraction.
The $R_1$ ratio in Fig.~\ref{fig:line_rat} has been normalised 
such that it tends asymptotically to 1.0
in the optically thin limit.

For times up to about 70 dy, 
$R_1$  
discriminates between different 
distributions of 
$^{56}$Ni with $\tau_C$ (see Fig.~\ref{fig:taus}).
Using this ratio, spherical models in which the radioactive material 
is behind most opacity 
(Model SS) and least opacity (Model SM) can be easily discriminated 
from the control model.
Aspherical models behave similarly -- in Model AO,
$R_1$ is largest 
if observed from the side to which the Ni blob is displaced;
the effect is present but weaker in Model AE because $\tau_C$ is
typically smaller.

As for $R_1$, for any
other pair of strong Co emission
lines the flux ratio $F(\epsilon_1)/F(\epsilon_2)$ will be determined
by the ratio of Compton opacity at the two line energies,
$\sigma_C(\epsilon_2)/\sigma_C(\epsilon_1)$. We will not discuss these
other potential diagnostic ratios but only comment that the most useful ratios will always be those between
lines of significantly different energy (as in $R_1$).
Thus, the ratio of the two strongest 
lines ($F($0.847~MeV$) / F($1.238~MeV$)$), although easier to measure, has 
less diagnostic value than $R_1$.

The second line-flux ratio shown in Fig.~\ref{fig:line_rat} is between
the $^{56}$Ni 0.158~MeV line and the $^{56}$Co 0.847 MeV line, referred
to as $R_{2}$.  In the figure, it is
normalised to the optically thin limit, a
decreasing function of time set by the decay rates of $^{56}$Ni
and $^{56}$Co.

\begin{figure*}
\epsfig{file=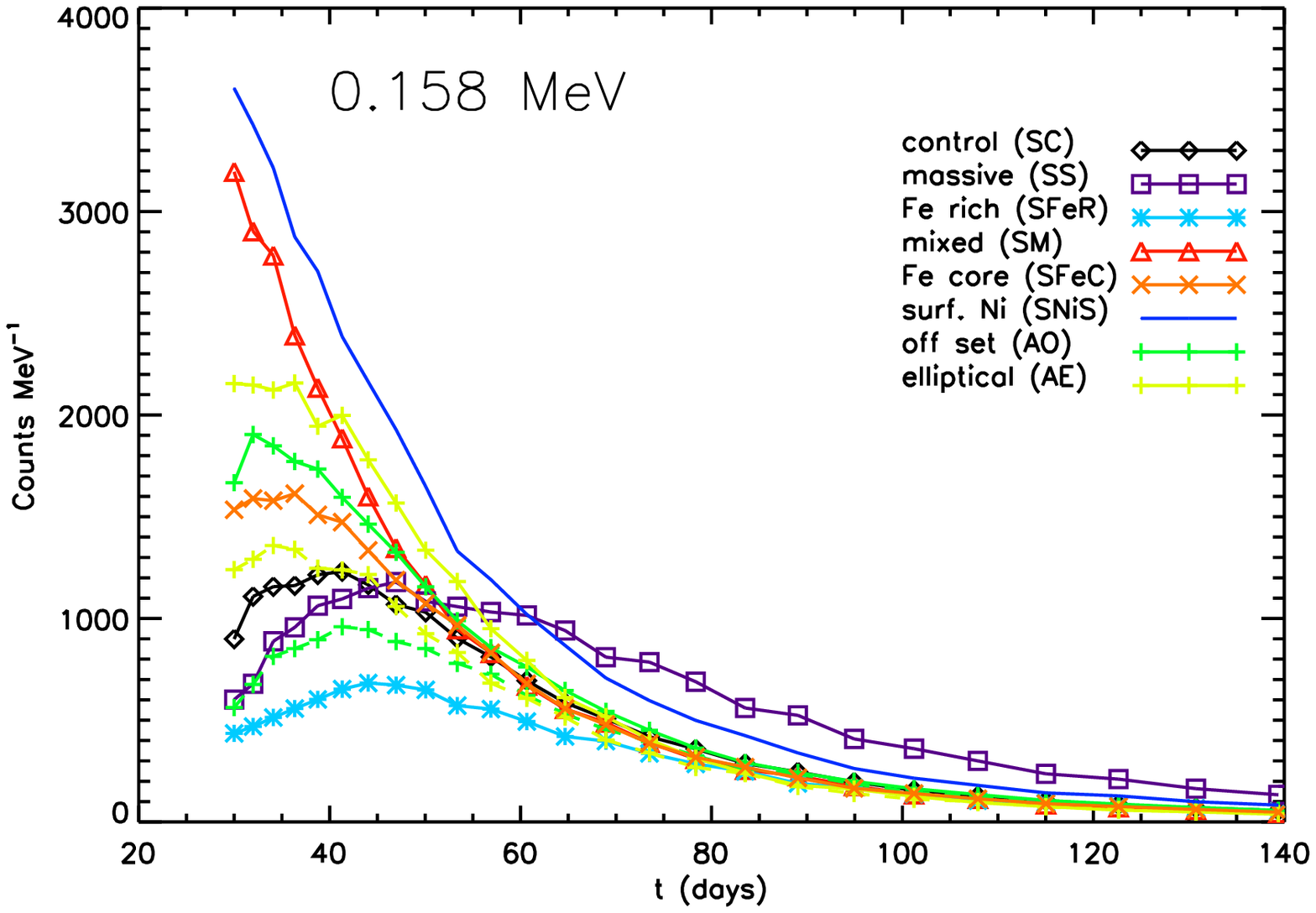, width=5.8cm}
\epsfig{file=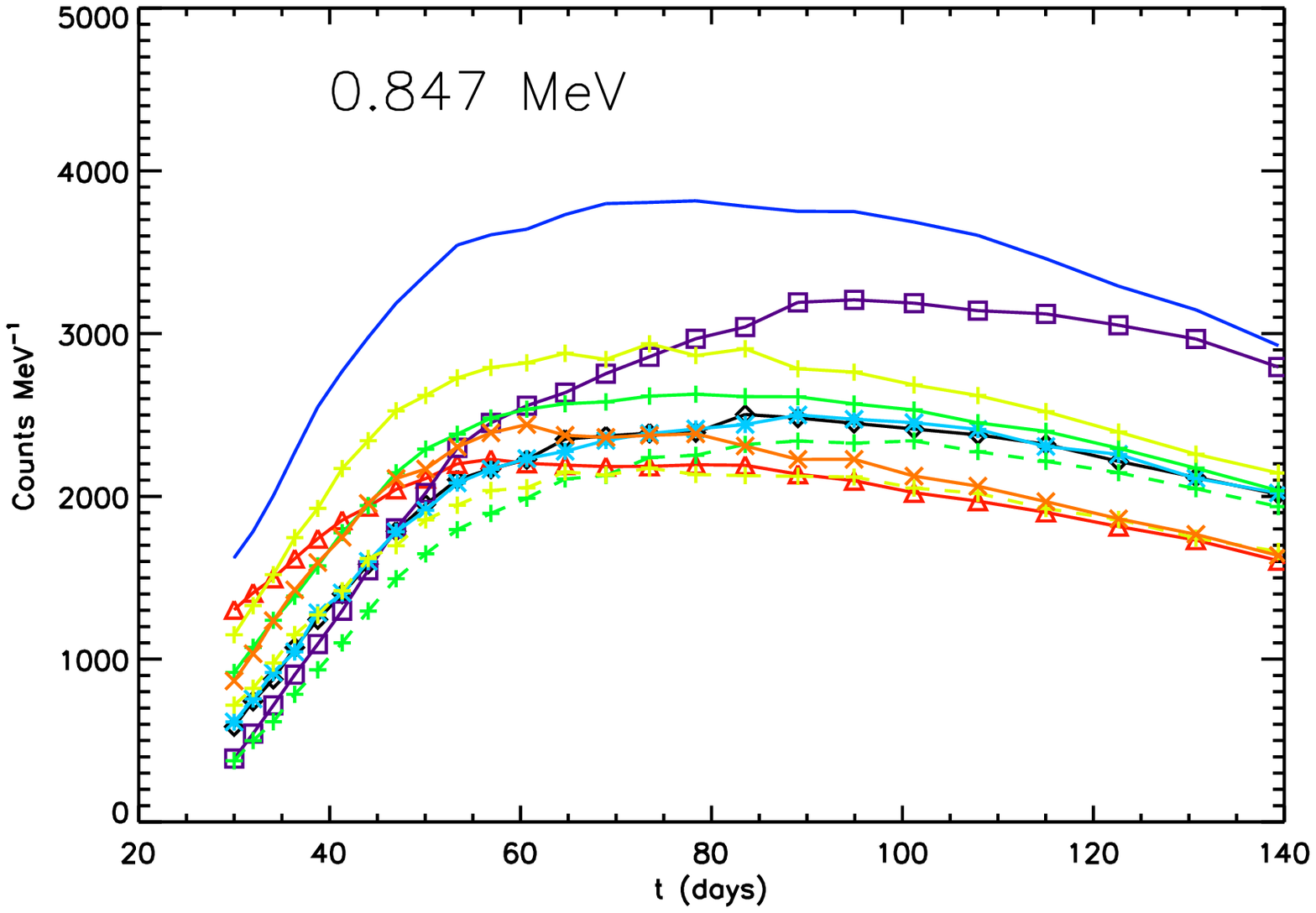, width=5.8cm}
\epsfig{file=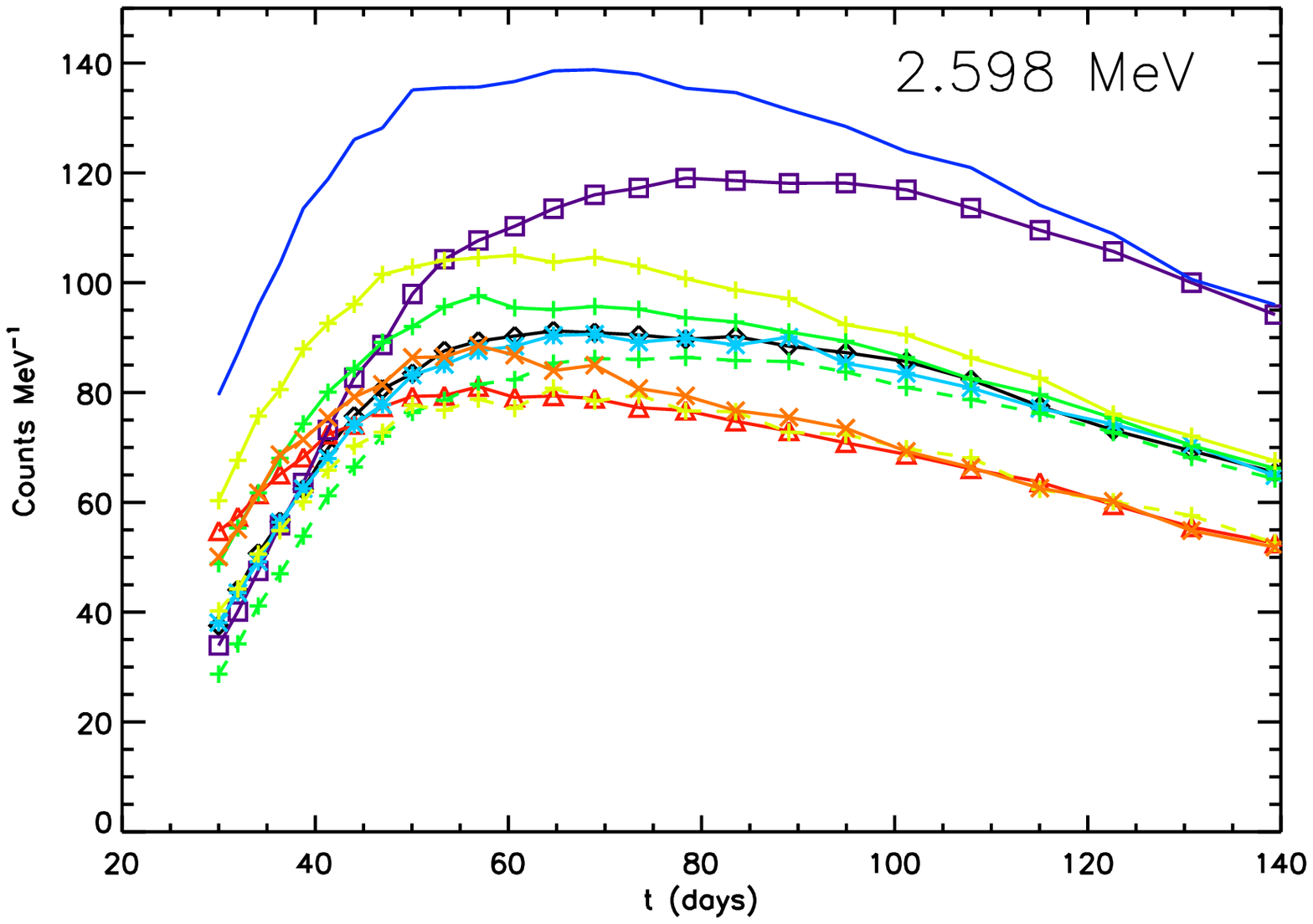, width=5.8cm}
\caption{
Predicted peak counts MeV$^{-1}$ for the 0.158, 0.847 and 2.598~MeV
emission lines computed 
adopting the {\it Integral SPI} effective area, a
recording interval of $10^5$~s and a source distance of 18~Mpc.
For
both the aspherical models (Models AO and AE), two curves are plotted 
-- as in Fig.~\ref{fig:taus}, these represent the
extremes with respect to viewing angle.
\label{fig:line-int}}
\end{figure*}

For most models, $R_2$ gives the same relative ordering
as $R_1$ at early times.
Again, this is primarily due to the
energy-dependence of $\sigma_C$. However, the 
effect is weaker in $R_2$ than $R_1$ because it is opposed by an
increase in the Compton continuum flux
around 0.158~MeV. 
Disentangling these two effects would require sufficiently high quality
data that 
the $\gamma$-ray continuum level around the 0.158~MeV line can be
measured.

Unlike $R_1$,
however, $R_2$ effectively separates Model~SFeR from any of the
others at $t < 50$~dy since the $0.158$~MeV line
is sufficiently soft to be affected by photoabsorption.
Thus, in agreement with \cite{gomez98}, we conclude that
$R_2$ (or an equivalent ratio of one of the
low-energy Ni lines to a harder Co line) provides an important 
diagnostic for composition.

\subsubsection{Diagnostic value}
\label{sect:line_rat_obs}

The $R_1$ and $R_2$ ratios defined above provide
useful diagnostics for two of the quantities that may be constrained
via the $\gamma$-ray spectrum, the distribution of total mass
and the composition of the photoabsorbing plasma.

Peak count rates for the relevant emission
lines computed adopting the same
conditions described in Section~\ref{sect:lc_obs} 
are shown in
Fig.~\ref{fig:line-int}.
For both the 0.158 and 0.847~MeV lines, 
the peak count rates are several thousand/MeV.
The intrinsic line widths, determined by the velocity
range in the ejecta, are $\sim 6$~keV for the 0.158~MeV line and
$\sim 34$~keV for the 0.847~MeV line. These widths are large enough to
be moderately well-resolved by modern instruments (\citealt{attie03,roques03}).
For our assumed distance and integration time, one would only
expect around 50 source counts in the 0.847~MeV line and fewer than 
10 source counts in the 
0.158~MeV line.

Owing to the combination of
fewer source photons and smaller adopted detector effective area, there are fewer
counts/MeV for the
2.598~MeV line -- only
a handful of counts integrated over the line profile.

Thus, statistical errors caused by small number counts are 
very substantial in both $R_{1}$
and $R_{2}$, meaning that these ratios could not contain useful information
for objects at our chosen distance. 
A tenfold increase in source counts would be needed 
to make the statistical accuracy
comparable to any of the
differences between the models (as shown in
Fig.~\ref{fig:line_rat}).
With 50 -- 100 times more counts and good time coverage 
these ratios could contain detailed information although additional
instrumental effects would further limit their value in practise. 

In light of this difficulty, in the next section we consider an alternative
approach in which 
the spectrum is coarsely binned and hardness ratios are extracted.

\subsection{Hardness ratios}
\label{sect:hard_rat}

\subsubsection{Discussion}
\label{sect:hard_rat_theory}

\begin{figure*}
\epsfig{file=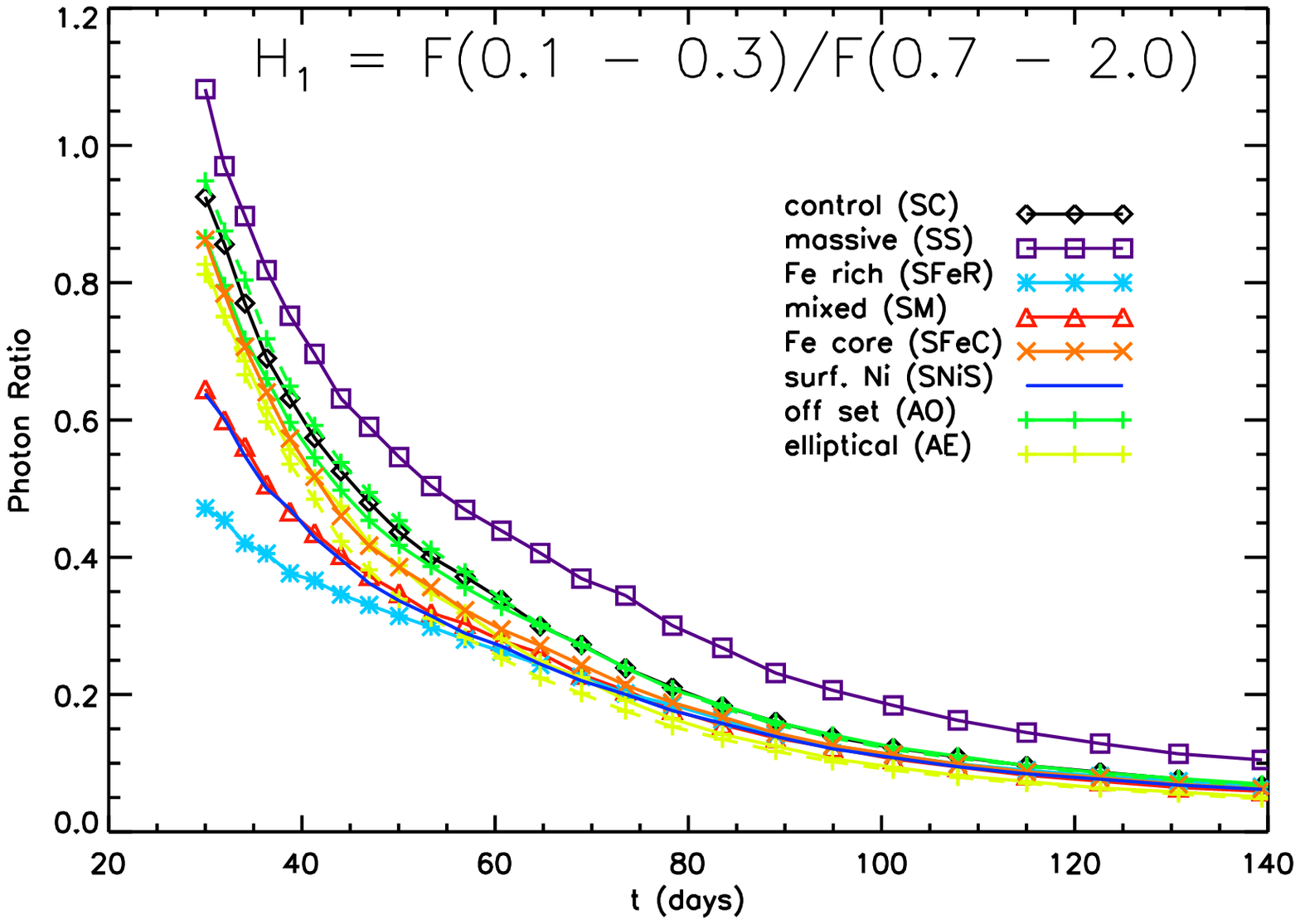, width=5.8cm}
\epsfig{file=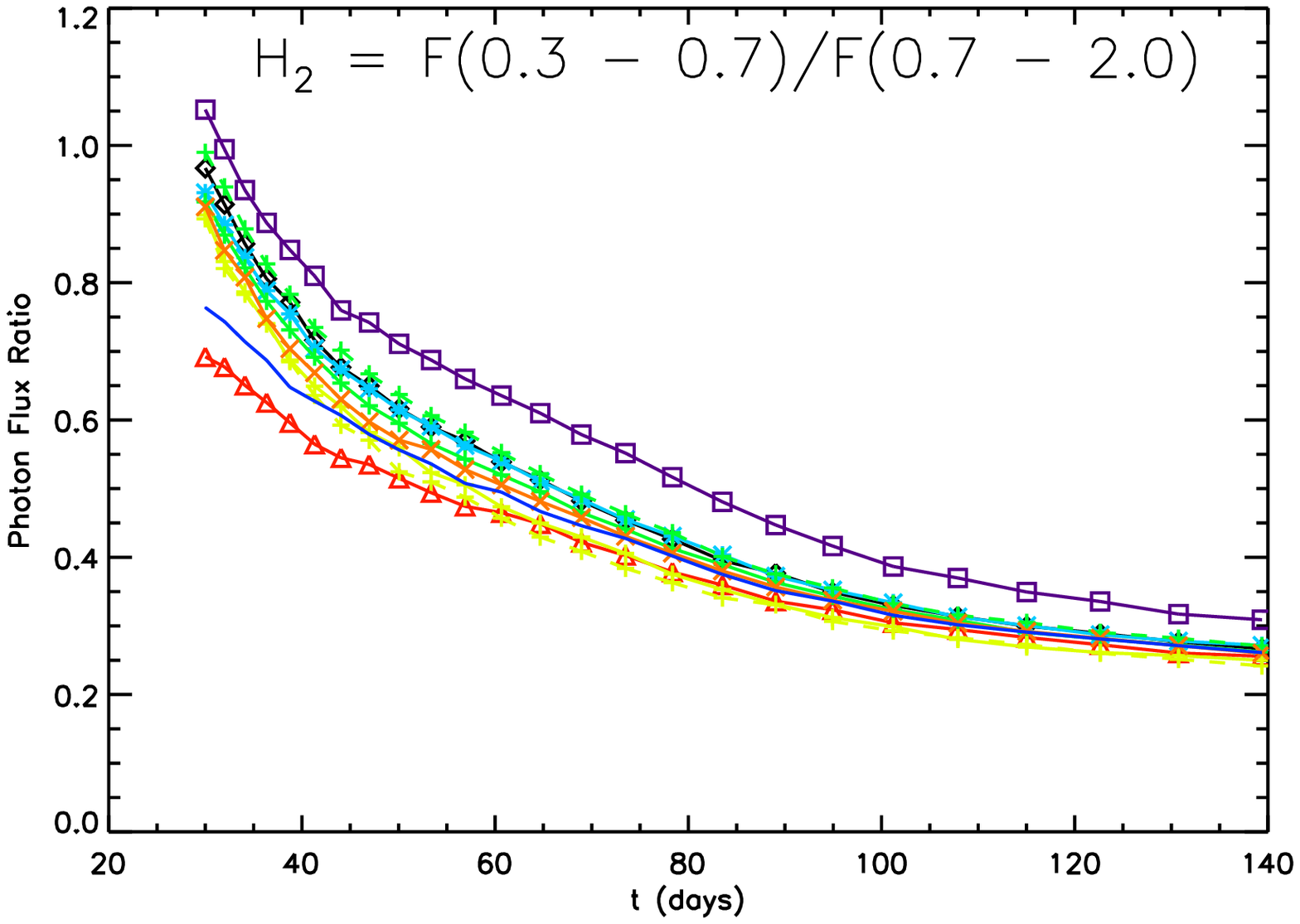, width=5.8cm}\\
\epsfig{file=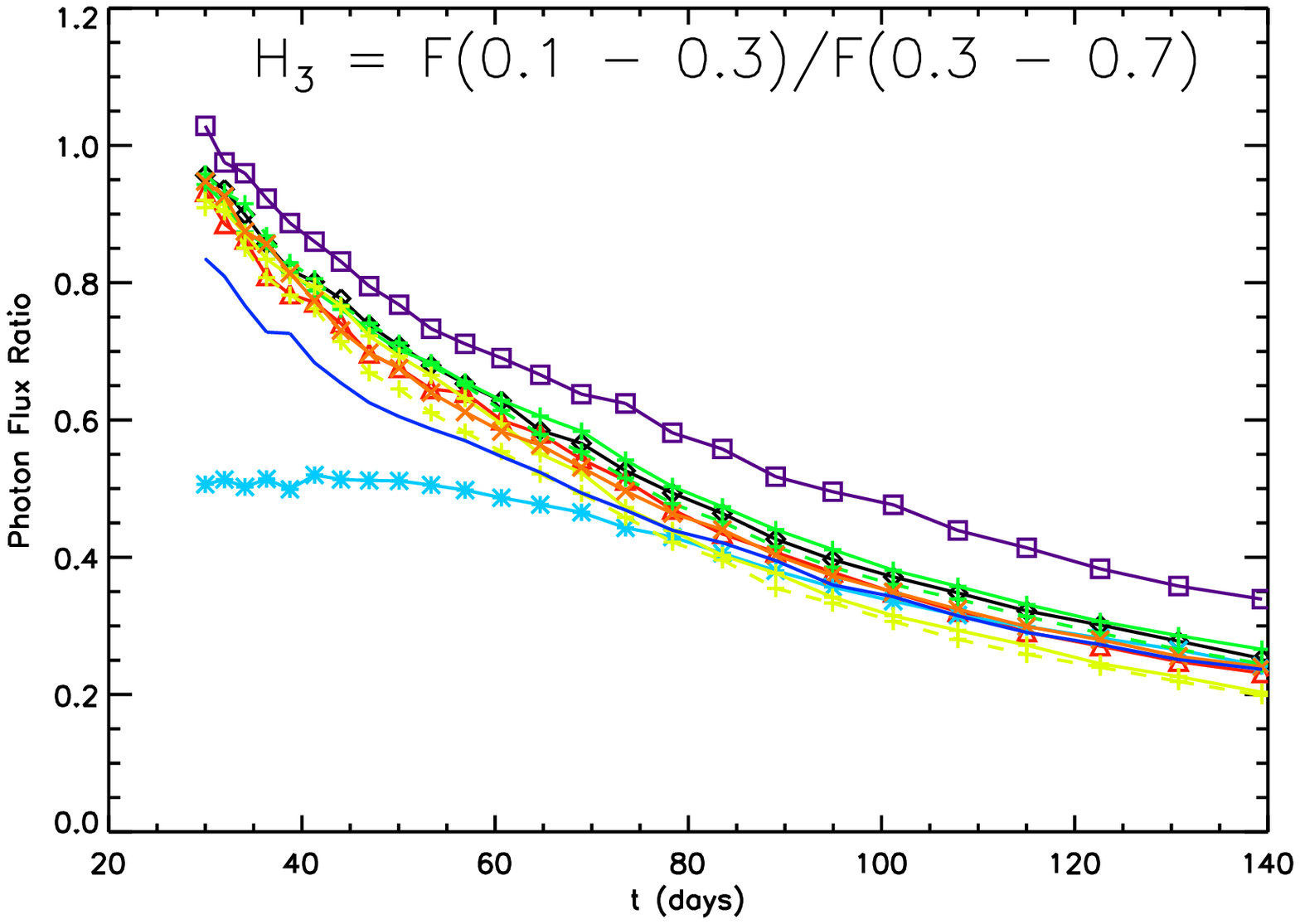, width=5.8cm}
\epsfig{file=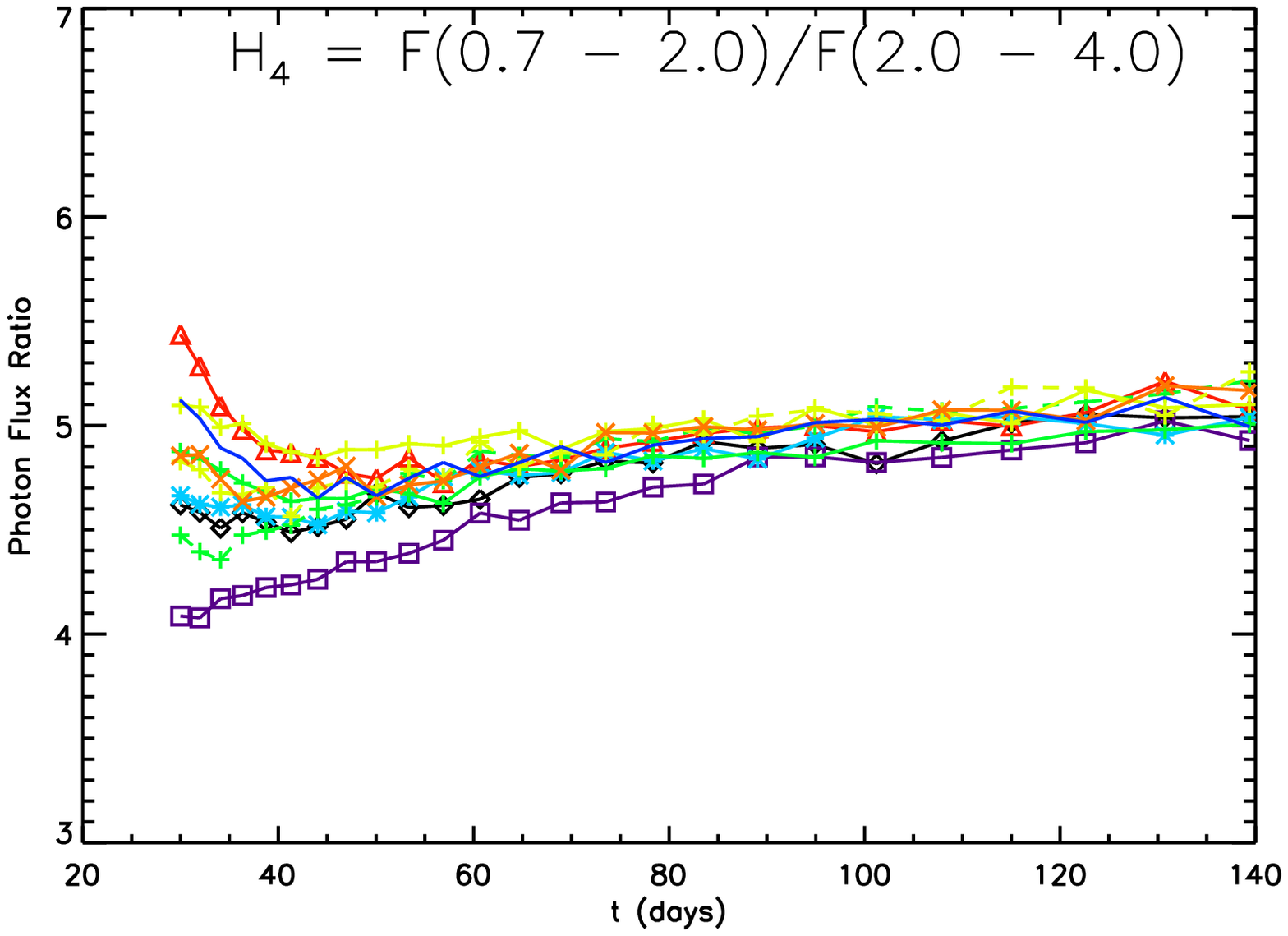, width=5.8cm}\\
\caption{
Hardness ratios
computed from the various models as functions of time since explosion. 
For the aspherical models, two curves are
plotted -- as in Fig.~\ref{fig:taus}, these represent the results for the two extreme observer
lines-of-sight.
\label{fig:hard_rat}
}
\end{figure*}

\begin{figure*}
\epsfig{file=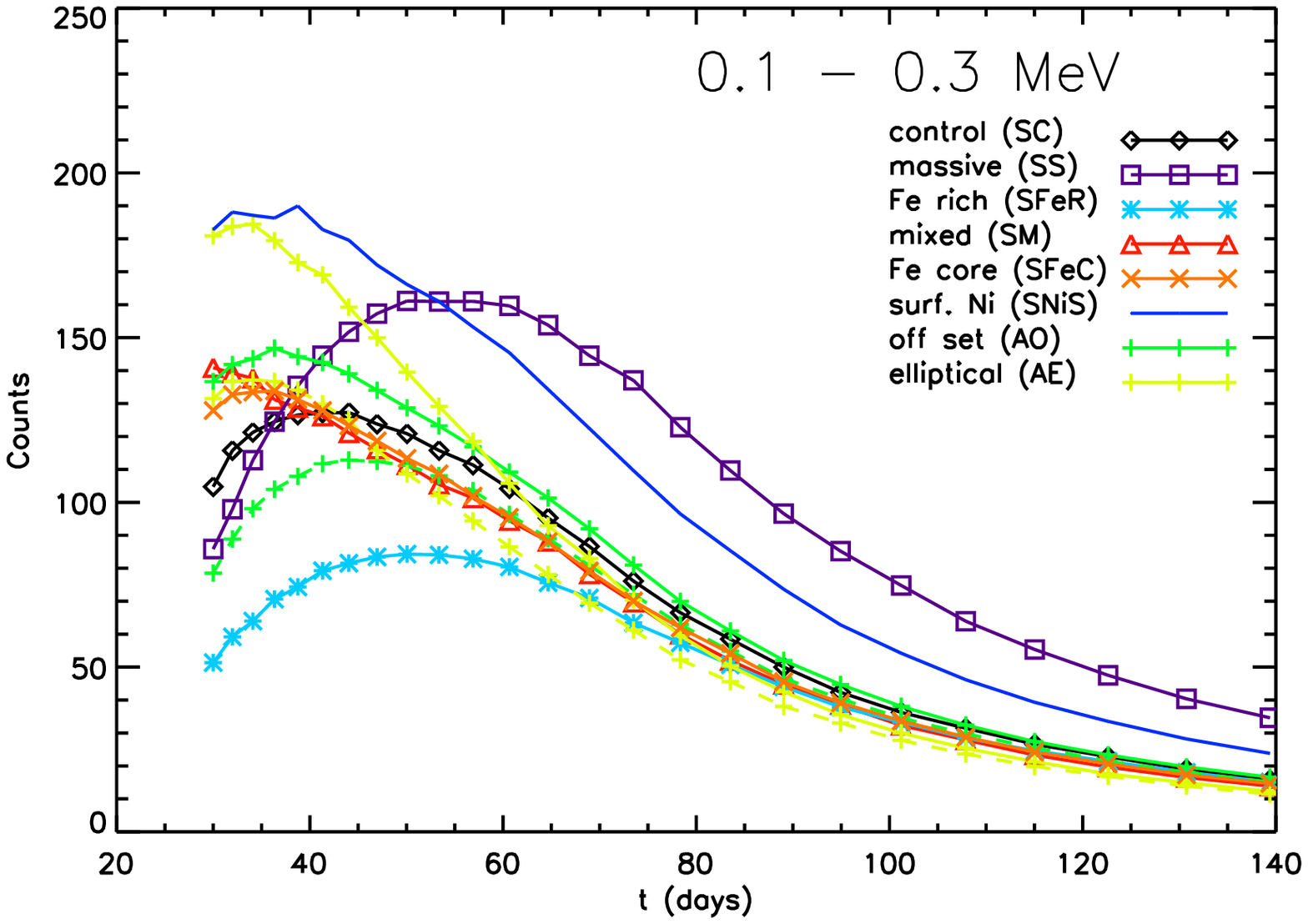, width=5.8cm}
\epsfig{file=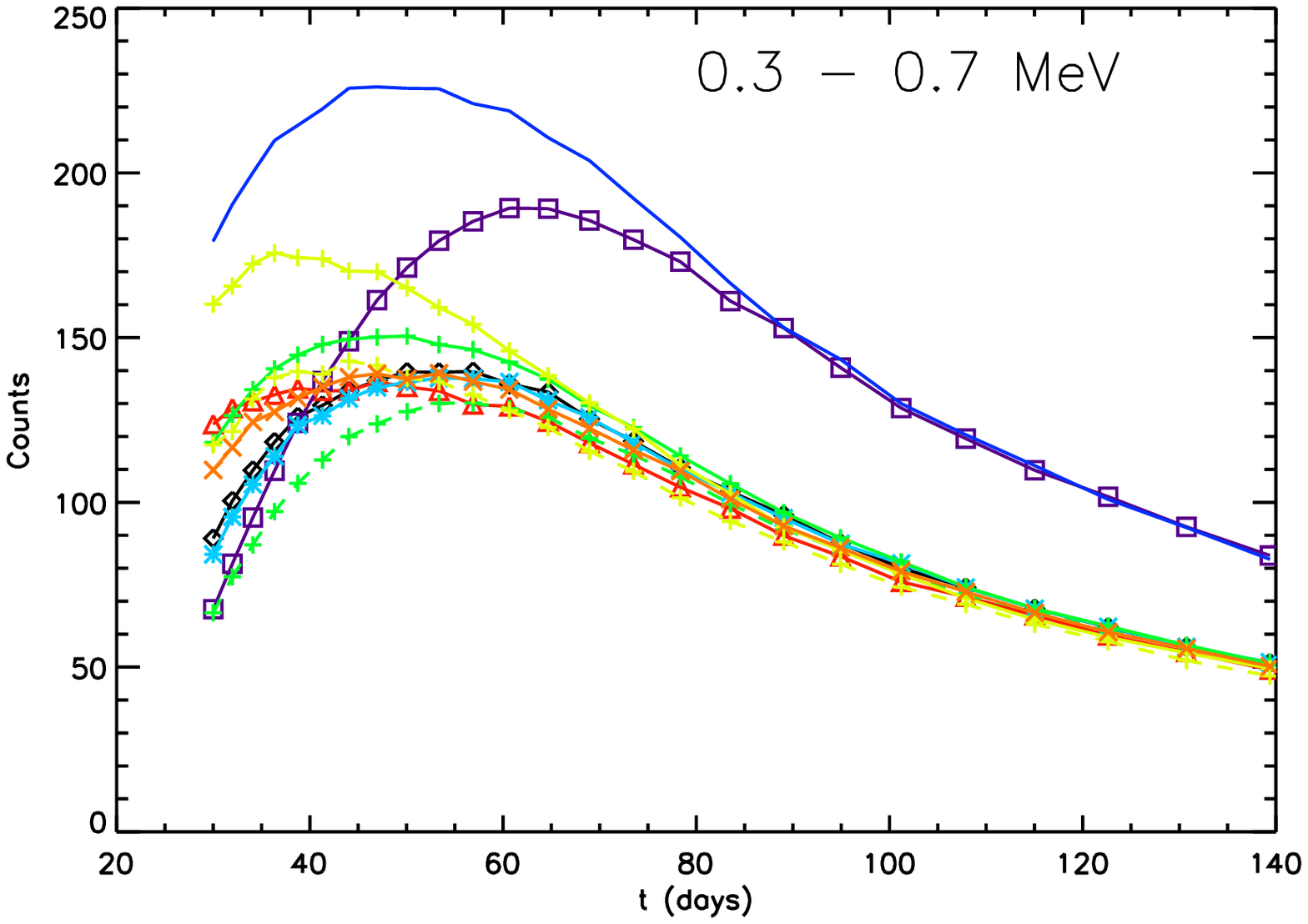, width=5.8cm}\\
\epsfig{file=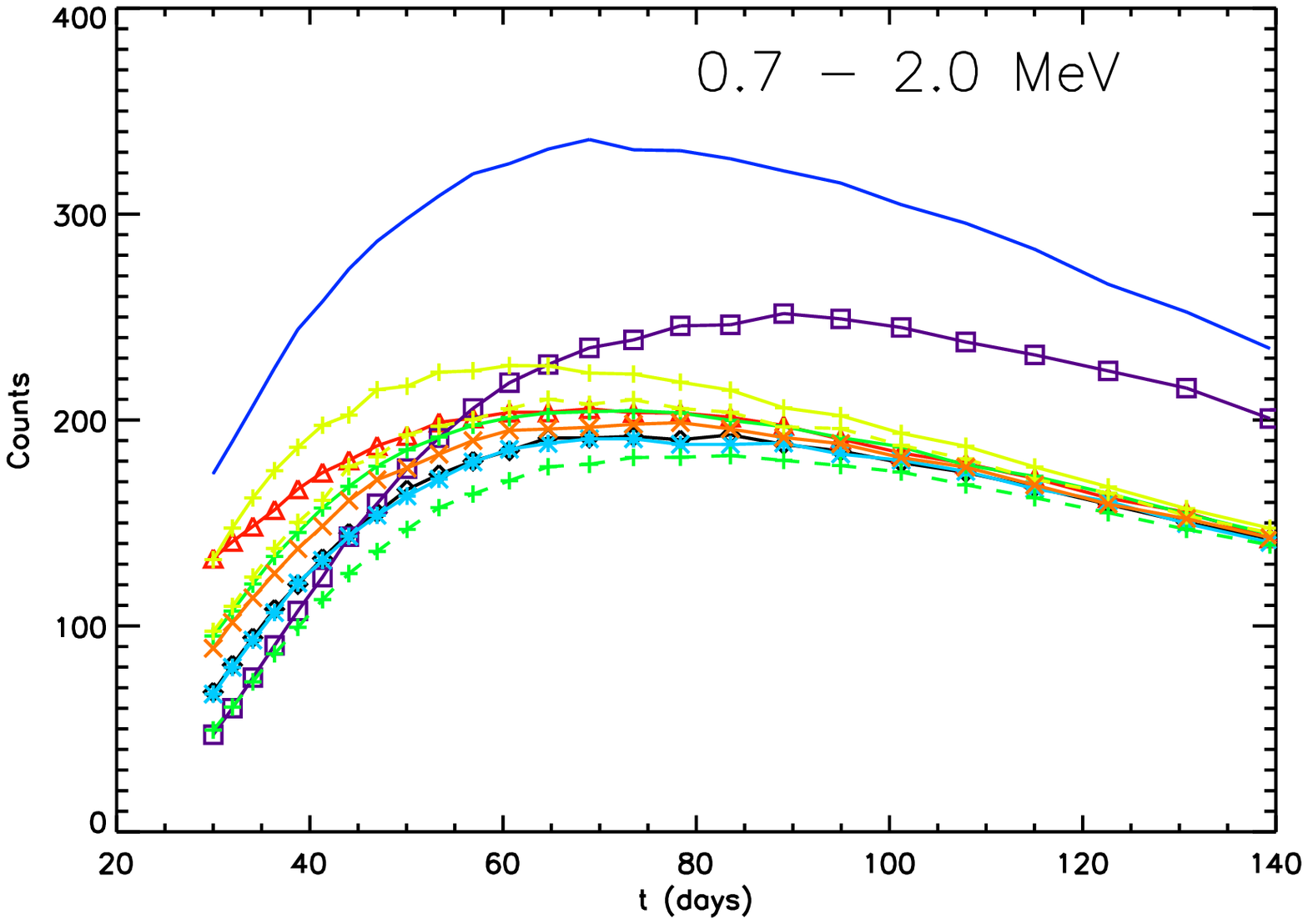, width=5.8cm}
\epsfig{file=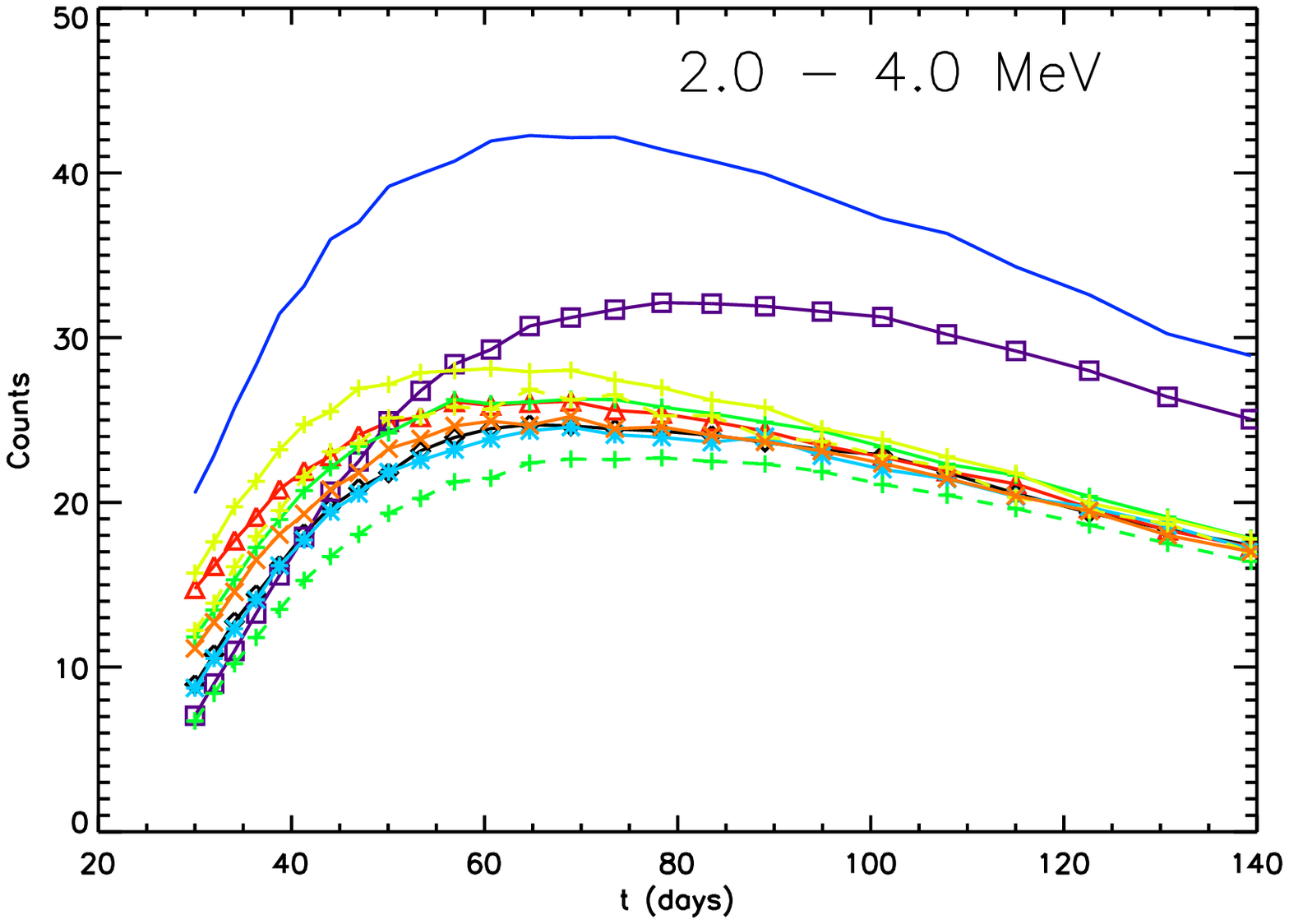, width=5.8cm}\\
\caption{
Synthetic light curves for the various models in each of the four
energy bands considered computed 
adopting the {\it Integral SPI} effective area, a
recording interval of $10^5$~s and a source distance of 18~Mpc.
For the aspherical models, two curves are
plotted -- as in Fig.~\ref{fig:taus}, these represent the results for the two extreme observer
lines-of-sight.
\label{fig:lcs}
}
\end{figure*}

Given the simplicity of
the $\gamma$-ray spectrum, moderately complete information can be
obtained from hardness ratios
alone; this is in contrast to other wavebands where
spectra are complex and not well-described by
photometry alone.
Some prospects of using relatively broad energy bands for
probing the soft, continuum regions of the spectrum were discussed by
\cite{gomez98}. We extend this to consider 
hardness ratios involving both continuum-dominated and 
higher energy line-dominated bands. 

We have divided the spectrum into four energy bands, two in
which the Compton continuum contributes significantly to the flux
($C_1$ and $C_2$) and two in which strong line
emission always dominates ($L_1$ and $L_2$). The bands are defined
below and are indicated in Fig.~\ref{fig:spec}:

\begin{itemize}

\item $C_1$ (0.1 -- 0.3 MeV): emission is
generally dominated by Compton down-scattering of harder photons, 
photoabsorption can be significant.

\item $C_2$ (0.3 -- 0.7 MeV): emission is predominantly through
Compton down-scattering. There
is some direct line emission, mostly via the 0.511~MeV
electron-positron annihilation line.

\item $L_1$ (0.7 -- 2 MeV): emission is dominated by strong lines
of Co supplemented by a moderately weak Compton continuum.

\item $L_2$ ($> 2$ MeV): lines of Co again dominate, now
with an even weaker Compton continuum.

\end{itemize}

The line ratios $R_1$ and $R_2$ (see Section~\ref{sect:line_rat}),
are affected by the change in $\sigma_C$ between different
energies. 
A similar
effect can be anticipated in the hardness ratio between the two line-dominated bands
($L_1$ and $L_2$) -- such an effect is present but
weak (see below). Another effect, however, can be noted when
comparing fluxes at soft energies. The strength of the
Compton continuum relative to the lines increases with  
opacity and thus high optical depths lead to softer hardness
ratios when $C_1$ or $C_2$ are involved.

This is illustrated in
Fig.~\ref{fig:hard_rat} which shows four hardness ratios computed with
different combinations of the four $\gamma$-ray bands. The hardness
ratios $H_{1} = C_1 / L_{1}$, $H_{2} = C_2 / L_{1}$ and $H_{3} = C_1 /
C_2$, all show the effect of high $\tau_C$ in
softening the spectrum. Of
these, $H_1$ and $H_2$ are best at
distinguishing the models, although the scale of the
differences is smaller than in the line ratios 
(c.f. Fig.~\ref{fig:line_rat}).

The hardness ratios involving the $C_1$ band ($H_1$ and $H_3$) are
sensitive to the composition of the scattering material -- these ratio
clearly separate Model~SFeR from Model~SC. This behaviour is
analogous to that of the $R_2$ line
ratio.

The fourth ratio, $H_4 = L_1 / L_2$ shows an
opposite trend  -- it
is hardest when optical depths are high (Fig.~\ref{fig:line_rat}). 
As mentioned above, this 
can be understood by 
analogy to $R_1$
but 
$H_4$ is less sensitive since it represents an average over a range of
energies.

Thus, $\gamma$-ray hardness ratios can contain quantitative information similar to that available
from the line ratios discussed in
Section~\ref{sect:line_rat}. Although hardness ratios are
less sensitive than line ratios, 
they may benefit from increased 
accuracy owing to the
broad energy ranges over which the spectra can be
integrated.

\subsubsection{Diagnostic Value}

Light curves of the various models 
in each of the four energy bands defined in 
Section~\ref{sect:hard_rat_theory} were computed, again adopting the same
conditions described in Section~\ref{sect:lc_obs}, 
and are shown in Fig.~\ref{fig:lcs}.

In the three softer bands, we expect roughly 100 -- 150 source counts
from most models at $\sim$ 40 dy
with the adopted observation parameters, giving
an optimistic estimate of $\sim$ 15 per cent statistical error in
$H_1$, $H_2$ and $H_3$. 
This is sufficient to separate 
the most different models -- Models~SC and SFeR could 
be distinguished via either $H_1$ or $H_3$ while Models SM and
SS would be measurably different from each other and marginally
distinguishable from Model~SC. However, there is little prospect of
using these ratios to find any of the effects of
asphericity captured by Models~AO and AE.

The predicted $L_2$-band source count 
rate is too low to allow the $H_4$-ratio to be useful --
only a handful of counts are anticipated in $L_2$ for any of
the models, a consequence of the reduction in both our adopted 
detector effective area and number of source photons at hard energies.

Still, hardness ratios could carry some diagnostic
information. 
In contrast to the line-ratio diagnostics 
(Section~\ref{sect:line_rat_obs}), an increase in the count number
by even a modest factor would allow fairly detailed analysis. 
An increase by a factor of $\sim$~10 -- say as a result of a serendipitously
nearby SN~Ia -- would reduce the source statistical error in the $H_1$, $H_2$ and
$H_3$ ratios to less than 10 per cent. At this level
of accuracy, these ratios would contain quantitatively useful information,
particularly if their time evolution could be followed 
through the peak of $\gamma$-ray emission.

\section{Summary and prospects}
\label{sect:summ}

Using a Monte Carlo code we computed
$\gamma$-ray spectra for a variety of models to explore whether unambiguous
constraints could be obtained from $\gamma$-ray observations of SNe~Ia.
Two aspherical 
toy geometries (a lop-sided distribution of Ni and an
ellipsoidal ejecta) show that moderate departures from sphericity
can produce viewing-angle effects at least as significant as
those due to 
variations of key parameters in 1D models. Thus $\gamma$-ray data could carry some
useful 
constraints on possible geometries, but caution must be
applied when evaluating its potential usefulness
in distinguishing specific explosion scenarios.

Given the limited sensitivity of current $\gamma$-ray missions, we
conclude, in
agreement with previous studies (e.g. \citealt{gomez98}), that there
is little prospect for obtaining useful constraints from line-ratio
diagnostics except for fortuitously nearby objects. Instead, we
suggest that the best 
prospects are offered by
hardness ratios. 
In particular, owing to the simplicity
of the physics underlying the $\gamma$-ray spectrum, a simple ratio of
the total emission in a hard energy, line-dominated part of the
spectrum to a soft energy, continuum-dominated region could
discriminate between our more extreme models. We also
emphasise the value of obtaining multiple observations over a wide
time period given the diagnostic power of 
the light curve shape. In planning future $\gamma$-ray missions, 
greater sensitivity at harder
energies ($\simgt 2$~MeV) 
should be given high priority
since 
several of the most potentially useful diagnostics
(e.g. our $R_{1}$- and $H_{4}$-ratios)
require measuring the hardest energies in the spectrum.  

\section*{Acknowledgments}

PAM thanks C. Wunderer and B. P. Schmidt, and 
SAS acknowledges R. Kotak and A. Watts, for stimulating discussions.
This research was supported in part by the National Science Foundation under 
Grant No. PHY05-51164.

\bibliographystyle{mn2e}
\bibliography{snoc}

\label{lastpage}

\end{document}